\documentclass[onecolumn,natbib,lscape]{aa}
\usepackage{natbib}
\usepackage{graphicx}
\usepackage{graphics}
\usepackage{longtable}
\usepackage{lscape}
\usepackage{amssymb}
\usepackage{txfonts}
\def\bb{} % per togliere il  grassetto dai cambiamenti
\def\arcsec{\ifmmode^{\prime\prime}\;\else$^{\prime\prime}\;$\fi}
\def\arcmin{\hbox{$^\prime$}}
\def\deg{\hbox{$^\circ$}}
\def\mJ{mJy beam$^{-1}$~}

\begin{document}
   \title{The Radio Luminosity Function of the NEP Distant Cluster Radio 
Galaxies}

   \author{M. Branchesi
           \inst{1,2}, 
           I. M. Gioia
           \inst{2} 
           C. Fanti
           \inst{2,3}, 
           R. Fanti
           \inst{2,3} and
           R. Perley
           \inst{4}
    }

   \offprints{M. Branchesi}

   \institute{Dipartimento di Astronomia, Universit\`a di Bologna,
              Via Ranzani 1, 40127 Bologna, Italy \\
                \email{m.branchesi@ira.inaf.it}
                           \and
            Istituto di Radioastronomia, INAF-CNR, Via Gobetti 101, 
            40129 Bologna, Italy \\
                \email{gioia@ira.inaf.it}
                          \and
           Dipartimento di Fisica, Universit\`a di Bologna, Via
           Irnerio 46, 40126 Bologna, Italy \\ 
                \email{rfanti@ira.inaf.it, cfanti@ira.inaf.it}
                          \and
           National Radio Astronomy Observatory, P.O. Box O,
                  Socorro, NM 87801, USA \\
           \email{rperley@nrao.edu}
             }

   \date{Received ......... 2005; accepted ......... 2005}

   \abstract{A complete sample of 18 X-ray selected clusters of galaxies
belonging to the {\em ROSAT} North Ecliptic Pole (NEP) survey has been 
observed  with the Very Large Array at 1.4 GHz. These are the most 
distant  clusters in the X-ray survey with redshift in the range  
0.3 $<$ z $<$ 0.8. Seventy-nine radio sources are detected within half an 
Abell  radius with an observed peak brightness $\geqslant$ 0.17 
mJy beam$^{-1}$, except for three sources, belonging to the same cluster, 
which have a higher peak brightness limit of 0.26 mJy beam$^{-1}$. The NEP 
field  source counts are in good agreement with the source 
counts of a comparison survey, the VLA-VIRMOS deep field survey, indicating 
that the NEP sample is statistically complete. Thirty-two out of the 79 
sources are within 0.2 Abell radii, twenty-two of them are considered cluster 
members  based on spectroscopic redshifts or  their optical magnitude and 
morphological classification.  The cluster radio galaxies  are used to 
construct the Radio Luminosity Function (RLF) of distant X-ray selected 
clusters. A comparison with two nearby cluster RLFs shows that the NEP RLF 
lies above the local ones, has a  steeper slope at low radio powers ($\le$ 
10$^{24}$ W Hz$^{-1}$) and shows no evidence for a break at  $\approx 6 
\times$ 10$^{24}$ W Hz$^{-1}$ which is observed in the nearby cluster RLFs. 
We discuss briefly the origin and possible explanations of the differences 
observed  in the radio properties of nearby and distant clusters of galaxies. 
The main result of this study is that the RLF of the distant X-ray clusters 
is very  different from that of the local  rich Abell clusters.

}
\authorrunning{Branchesi et al.}
\titlerunning{RLF of NEP X-ray  Galaxy Clusters}

\maketitle
\section{Introduction}

One of the most fascinating topics in cluster astrophysics is the study of 
the effects produced by cluster formation on the properties of the radio 
sources  embedded in the Intra Cluster Medium (ICM) and on the evolution of the
cluster radio galaxies. Since the pioneering work of \cite{hl91} \citep[see 
also][]{eyg91}, who showed that powerful classical radio sources at high 
redshift (z $\sim$ 0.5) preferentially inhabit rich environments, several  
other authors have investigated the link between cluster environment and radio
source morphology, or between cluster formation and  radio galaxy triggering,
or cluster mergers and star formation triggering  \citep[see among others:]
[]{oo85, bes02, sr02, ol99, owen05}. Extremely distant cluster environments, 
or proto-clusters, have been found by targeting  objects near powerful radio 
sources  \citep{cf96, pen97, pen99}  with a  redshift up to z $\sim$ 4 
\citep{ven02}. Other lines of investigation, dating back to the 1970s 
\citep{or76, bor79} involve the morphology and radio power  of nearby 
cluster radio sources  \citep{ol89, ow91}.

\medskip
\noindent
A different approach  from those studies which try to understand the 
environment  effects by studying individual sources  \citep[i.e.][]{pr99,
ven02} or individual  clusters \citep[i.e.][]{ol99,owen05}, would  be to start 
from a  well defined sample of clusters of galaxies and to obtain moderately 
deep radio  observations  to investigate the properties of radio sources in 
clusters  and their evolution. Exhaustive and in depth studies of 
optical/radio  properties of radio galaxies in nearby (z $\le$ 0.25)  Abell  
clusters  have appeared in the literature \citep[over 500 Abell clusters
studied  with the VLA by Owen, Ledlow \& collaborators; see][ and references 
therein]{ol97}. {\bb For more recent work on the subject see \cite{mo03}.}
An equivalent study for distant clusters (i.e.
up to z$=$0.7--0.8) is still missing even if radio surveys of a small 
number of high-z, rich, X-ray selected  clusters have  been conducted 
\citep[e.g.][]{sto99, per03} mainly with the  goal of investigating 
the evolution of the cluster radio galaxy population. According to these 
authors an apparent dearth of radio sources below log P (W Hz$^{-1}$) $=$ 24.5
is due to incompleteness rather than to a weak negative evolution. New, 
deeper radio observations would discriminate on the reality of this 
possible evolution in the cluster radio population. Also, better optical 
identifications of radio sources with  cluster members would help in 
addressing the issue of evolution. 

\medskip
\noindent
We present in this paper the results from a VLA survey which uses a new sample 
of X-ray selected, distant clusters of galaxies extracted from the {\it ROSAT}
North Ecliptic Pole (NEP) survey  \citep{gio03}.  The goals are to provide an 
additional sample of distant objects observed in radio to be compared with 
the detailed studies of radio galaxies in nearby clusters, and to relate the 
radio properties of the cluster galaxies with the X-ray  and optical 
properties of the  clusters. Searching for differences in the radio luminosity
function between  local and distant galaxy clusters will provide a  valuable
perspective on the evolution of radio galaxies, and on the influence
of the environment on radio  galaxy populations.

\medskip
\noindent
Throughout the paper we assume H$_0=$ 75 km s$^{-1}$ Mpc$^{-1}$ and 
q$_0 =$ 0.1. Even though we are aware that this is not the currently-favored 
cosmological concordance model, the cosmology we adopted  allows us to make a 
direct  comparison  to previous work in the field. As explained in 
Section \ref{RLF-comp}, the results obtained here would be strengthened if 
one adopts the concordance  cosmology.

%________________________________________________________________
\section{The X-ray Selected Cluster Sample}

A complete sample of 18 X-ray selected galaxy clusters  was extracted 
from the {\em ROSAT} NEP survey catalogue \citep{gio03} for observations 
with the VLA at 1.4 GHz.  The sample contains the most distant clusters 
in the survey with redshifts in the range 0.3$<$z$<$0.8. The NEP survey 
\citep{mul01,hen01,vog01} covers a 87.4 deg$^{2}$ contiguous region of the 
{\em ROSAT} All-Sky Survey (RASS) \citep{vog99}, at a moderate Galactic 
latitude of  $b=29.8^{\circ}$, around  the North Ecliptic Pole 
($\alpha_{2000} = 18^{h}00^{m}$, $\delta_{2000} = $+$66^{\circ}33^{'}$).
The region around the NEP possesses the deepest exposure and consequently the
greatest sensitivity of the entire RASS. Hence, the 9\deg$\times$ 9\deg
survey region covers the deepest, wide-angle contiguous region ever observed 
in X-rays. This unique combination of depth plus wide, contiguous solid angle 
provides the capabilities of detecting both high-redshift objects and large 
scale structure. The survey catalogue contains 445 X-ray sources above a flux 
of $\sim$2$\times 10^{-14}$ ergs cm$^{-2}$ s$^{-1}$ in the 0.5--2.0 KeV 
energy band. A comprehensive program of optical follow-up observations to
determine the nature of each of the X-ray sources in the NEP sample led
to the identification of 63 clusters of galaxies. The identification of a
X-ray source as a cluster of  galaxies usually requires the absence of 
emission lines, the absence of a non-thermal  continuum, which 
identifies  a BL Lac object, coupled with a centrally 
concentrated galaxy overdensity either from  the POSS or from deep optical 
CCD images taken for more distant clusters, and at least two  concordant 
galaxy redshifts. The optical identification  spectroscopy was done 
using the University of Hawai$'$i (UH) 2.2m Wide-Field Grism Spectrograph 
(WFGS), the Multi-Object-Spectrograph (MOS) on the Canada-France-Hawai$'$i 
3.6m  telescope, and the Low-Resolution Imaging Spectrograph (LRIS) on Keck. A 
detailed description of the X-ray sources and their optical identifications 
can be found in \cite{mul01} and  \cite{gio03}. The 18 X-ray selected 
clusters and their basic parameters are listed in Table~\ref{tab1}. The 
columns contain the following information:
%-------------------------------------------------------------------------
\begin{itemize}
\item Column~1: Source name formed by the acronym RX\,J ({\em ROSAT} X-ray 
              source, Julian 2000 position),  and the X-ray centroid position
\item Column~2: Internal NEP source identification number
\item Column~3-4: Right ascension and declination of the optical Brightest 
                Cluster Galaxy (BCG) {\bb from the DSS\,II}
                (J2000, HH MM SS.S, $+$DD MM SS) 
\item Column~5: Total unabsorbed flux in the 0.5-2.0 KeV band in units
                of  $10^{-14}$ erg cm$^{-2}$ s$^{-1}$ 
\item Column~6: Spectroscopic redshift
\item Column~7: Abell radius (R$_{A} =$ 2 Mpc) in arcmin
\item Column~8: Rest frame {\it K}-corrected X-ray luminosity in the 
                0.5-2.0 KeV band in units of  $10^{44}$ erg s$^{-1}$
\item Column~9: Apparent red magnitude of the BCG from the SuperCosmos 
                POSS\,II (except for three cases indicated by an asterisk: for 
                RXJ\,1753.3$+$6631 and RXJ\,1821.6$+$6827 no magnitudes are 
                available while for RX\,J1751.5$+$7013  only the  POSS\,I 
                magnitude is available).
\end{itemize}

\section{Radio Observations and Data Reduction}
\label{radio-obs}

Observations of the 18 X-ray clusters were carried out at 1.4 GHz, with two
50 MHz IF pairs, on 2003, November 30 and December 7, using the Very 
Large Array (VLA) in B-configuration for a total integration time of 20 hours.
The B configuration was optimal to map extended areas with an acceptable
nominal resolution of 3.9\arcsec in both coordinates ~down to a faint 
brightness limit ($\approx$ 0.17 mJy beam$^{-1}$, see below). The 18 galaxy 
clusters  were observed in continuum mode, using the full available bandwidth 
per IF.  For each cluster one hour integration, divided into three  
"snapshot'' scans, was obtained by pointing the array at the position of the 
brightest central galaxy, or BCG, listed in Table~\ref{tab1}.  A typical plot 
of the  $u-v$ coverage is  shown in Fig.~\ref{fig1}.

%______________ Table 1 ________________________________________________
\begin{table}[htb]
\begin{center}
\caption{The X-ray selected cluster sample}
\begin{tabular}{lcccccccll}
\hline 
\hline
Cluster Name & NEP & RA  &  DEC  & $f_{X}$ & z & R$_A$ & L$_{X}$ & BCG& \\
~ & Id \# & J2000 & J2000 & 10$^{-14} cgs$ &~ & \arcmin & 10$^{44} cgs$ & ~~~m$_{R}$ &\\
\hline
RX\,J1753.3$+$6631&  200& 17 57 19.4&  $+$66 31 31&  ~3.87& 0.691 & 5.51 & 0.50& ----- *& \\   
RX\,J1758.9$+$6520&  310& 17 58 56.5&  $+$65 21 05&  ~4.53& 0.365 & 7.58 & 0.14& 20.1&\\
RX\,J1727.4$+$7035& 1730& 17 27 33.6&  $+$70 35 47& 50.20& 0.306 & 8.46 & 1.02&  18.4&\\
RX\,J1728.6$+$7041& 1780& 17 28 38.2&  $+$70 41 03& 28.26& 0.551 & 6.08 & 2.01&  19.9&\\
RX\,J1743.4$+$6341& 2420& 17 43 30.4&  $+$63 41 41& 50.49& 0.327 & 8.11 & 1.18&  17.7&\\
RX\,J1745.2$+$6556& 2560& 17 45 18.2&  $+$65 55 42&  ~7.09& 0.608 & 5.81 & 0.67& 19.7&\\
RX\,J1746.7$+$6639& 2770& 17 46 46.8&  $+$66 39 04&  ~8.81& 0.386 & 7.33 & 0.31& 18.5&\\
RX\,J1747.5$+$6343& 2870& 17 47 31.1&  $+$63 45 23& 15.71& 0.328 & 8.09 & 0.38&  18.3&\\
RX\,J1748.6$+$7020& 2950& 17 48 39.1&  $+$70 20 42& 13.84& 0.345 & 7.84 & 0.38&  18.6&\\
RX\,J1749.0$+$7014& 2980& 17 49 04.5&  $+$70 14 45& 23.79& 0.579 & 5.94 & 1.90&  18.0&\\
RX\,J1751.5$+$7013& 3130& 17 51 32.6&  $+$70 13 22& 13.10& 0.493 & 6.42 & 0.76&  19.8*&\\
RX\,J1752.2$+$6522& 3200& 17 52 08.2&  $+$65 22 53&  ~6.44& 0.392 & 7.27 & 0.24& 17.2&\\
RX\,J1754.5$+$6904& 3320& 17 54 34.2&  $+$69 05 07&  ~7.39& 0.511 & 6.30 & 0.48& 19.7&\\  
RX\,J1755.9$+$6314& 3450& 17 55 57.8&  $+$63 14 09& 17.92& 0.385 & 7.34 & 0.61&  18.5&\\
RX\,J1806.1$+$6813& 4150& 18 06 04.8&  $+$68 13 16& 19.91& 0.303 & 8.51 & 0.41&  19.0&\\
RX\,J1811.3$+$6447& 4560& 18 11 19.3&  $+$64 47 23& 13.53& 0.451 & 6.72 & 0.65&  20.1&\\
RX\,J1812.1$+$6447& 4610& 18 12 08.2&  $+$63 53 32& 17.25& 0.541 & 6.13 & 1.20&  20.4&\\
RX\,J1821.6$+$6827& 5281& 18 21 32.9&  $+$68 27 55& 10.22& 0.811 & 5.20 & 1.76&  ----- *&\\
\hline
\label{tab1}
\end{tabular}
\end{center}
\end{table}
%______________________________________________________________
%----------------------------------------------------------------------
%Fig.1
\begin{figure}
\centering 
\vskip 2truecm
\caption{Typical $uv$ coverage of the NEP radio survey. {\em This figure can 
be  asked directly to the first author}}
\label{fig1}
\end{figure}
%-------------------------------------------------------------------------

\medskip
\noindent
Our primary flux density calibrator was  3C286, whose flux density at 1.4
GHz was 14.9 Jy. The phase calibrators were 1748$+$700, 1634$+$627, 1849$+$670, 
which were given 2 minutes integration time at approximately 40 minute 
intervals to monitor instrumental and atmospheric phase and gain variations.
The data were reduced using the NRAO AIPS (Astronomical Image Processing 
System) package, following the standard procedure: calibration, Fourier 
inversion with a uniform  uv-weight, clean and restore. For each 
pointing an area of 1024 $\times$ 1024 pixels centered on the optical 
BCG was imaged. Given the size of each pixel  (1 pixel $=$ 1.5$''$) 
each image measures 25.6 $\times$ 25.6 arcmin. 

\medskip
\noindent
Strong radio sources far away from the field center, but within the
primary field of view, have sidelobes that might contaminate  the inner 
portion of the field.  To keep under control the confusion effects from 
these sources a multi-field reduction was applied. The NVSS (NRAO VLA Sky 
Survey, \citealt{con98}) catalogue was used to select all the sources with 
an attenuated peak brightness higher than 5 mJy beam$^{-1}$ within a radius 
of 30 arcmin from each pointing and not included in the 1024 $\times$ 1024 
pixel region (even though 2048 $\times$ 2048 pixel dirty images were created  
to  check that no additional confusing sources, other than the NVSS sources, 
were  present). For each field an AIPS ``{\it RUN}  file'' was produced to
define extra fields (64 $\times$ 64 pixels in size) containing these 
confusing sources. Several iterations of self  calibration and imaging 
were simultaneously applied to the main field  and to these additional 
satellite  fields.  At the beginning of each iteration the cleaning 
procedure was guided by setting boxes around the stronger sources of the 
central field and adding progressively other boxes around the fainter sources 
which were becoming visible. At the end a general cleaning was applied.
The adopted clean gain was 0.05. The clean was generally halted when 
the residuals reached a flux density comparable to three times the nominal 
noise level. The typical number of independent components searched  in 
the main field was about ten thousand. The a--priori calibration uncertainties
are $\epsilon=0.03$ for flux densities and $\epsilon=0.02$ for diameters.
The $r.m.s.$ noise in the final maps is in the range 0.026--0.034 \mJ  
(see Table~\ref{tab2}, column 2). Exceptions are RX\,J1751.5$+$7013 
(NEP\,3130) and RX\,J1751.5$+$7014  (NEP\,2980) where the  presence of 
very strong confusing sources, far from the field center, degraded the 
quality of the maps. The high level of the noise (1 \mJ) in the 
RX\,J1751.5$+$7014 field  prevented detection of any source. For this 
reason NEP\,2980 is not listed  in Table~\ref{tab2} and is not used in any 
subsequent analysis.

\medskip
\noindent
Radio sources with peak brightness $\ge$ 0.17 \mJ  have been  systematically 
searched over the whole area within half an Abell radius (0.5 R$_{A}$), 
namely within 1 Mpc in the cosmology adopted here, in all clusters except for 
NEP\,3130.  A peak brightness of 0.17 \mJ corresponds to five times the 
highest $r.m.s.$ of all maps (except for the two higher noise maps). 
The limiting  brightness of 0.17 mJy beam$^{-1}$ corresponds to a limit in the 
radio power of P$_{1.4 GHz}=$ $3.6\times10^{22}$ W Hz$^{-1}$ for the nearest 
cluster  (z$=$0.303) and of  P$_{1.4 GHz}=3.4\times10^{23}$ W Hz$^{-1}$ for 
the most distant one (z$=$0.811). Since  one half Abell radius corresponds to 
$\sim 4$ arcmin for the  closest clusters (or less  for the more  distant 
ones) the corrections to the flux densities for the primary beam attenuation 
are negligible.  A more serious effect is produced by the bandwidth smearing 
which depresses  the peak brightness of the sources (thus lowering the 
sensitivity at the  map periphery) and radially distorts them as a function 
of the distance  from the field center (see Appendix, Section~\ref{bs}). 
The peak response of a point source declines while the total flux density is
conserved. All peak brightnesses are corrected for this effect (see Appendix,
 Section~\ref{OPTS}).

%-----------------------------------------------------------------
\section{The Sample of Radio Sources}
\label{sample}

The final radio source sample contains 79 sources within half an Abell 
radius  with an observed peak brightness   S$_{P}\geqslant$ 0.17 mJy 
beam$^{-1}$ except for the three sources in NEP\,3130, which have a peak 
brightness  S$_{P}\geqslant$ 0.26 mJy beam$^{-1}$. A more detailed description
of the  data analysis is  given in the Appendix. Here we mention that
the source parameters listed below were obtained by a Gaussian
fit, when acceptable, using the AIPS task {\it JMFIT}. The parameters
are: position, maximum and minimum angular size,  both observed  
($\theta_M, \theta_m$) and deconvolved ($\Theta_M, \Theta_m$) for the 
map beam, peak brightness (S$_{P}$)  and total flux density (S$_{T}$)  
of the sources. Following  \cite{con98} we considered  as 
resolved in each direction those sources for which the fitted sizes, 
$\theta_{M,m}$,  exceed the beam Full Width Half Maximum (FWHM) by more 
than 2.33 times the estimated  errors.  Since the probability that this is 
due to statistical errors is  $<$~2\%, only $\sim$ 1 source would be 
mistakenly classified as slightly resolved instead of point--like.   
Since images appear convolved with a position--dependent, radially 
elongated beam due to smearing, we adopted the FWHM of the synthesized  beam,  
$\theta_b$, for the minor axis and the smeared FWHM $\theta_{\rm  bws}$ 
(see eq.~\ref{b_smea}) for the major axis, taking into account 
also the direction of the elongation.

\smallskip
\noindent
When the source structure was too extended to be reliably fitted by a 
multi-gaussian model, the peak and total flux densities were 
computed using  the AIPS task {\it TVSTAT}, which allows integration of 
the image brightness values over irregular areas. Sizes were derived from 
the lowest reliable contour in the image map. 
Both peak brightness and total flux density are 
optimized by accounting for noise and algorithm effects on the direct 
parameter determinations (see Appendix, Section~\ref{OPTS}). Parameter 
errors are derived  as in  Appendix (Section~\ref{errors}). 

\subsection{The Radio Source Catalogue}  

The 79 radio sources detected within half an Abell radius and their properties 
are listed in Table~\ref{tab2}. The columns contain the following information:

%--------------------------------------------------------------------
\begin{itemize}
\item Column 1: Cluster name
\item Column 2: Average image noise within half an Abell radius in \mJ
\item Column 3: Source identification letter
\item Column 4-5: Source radio position - Right Ascension and Declination 
              (J2000). For sources with a multiple structure 
           the coordinates refer to the radio centroid while for all the 
           others, including complex morphology sources, the coordinates 
           refer to the position of the radio peak. Exception is source
           NEP\,2420 D where the radio position refers to the midpoint
           between the two brightest peaks  
\item Column 6: Distance from the cluster center in units of Abell
             radius defined as R/R$_{A}$ 
\item Column 7-8: Bandwidth smearing corrected peak brightness and 
             integrated flux density with respective errors according to 
             the Appendix.  For sources too extended for a multi--gaussian fit
             the integrated flux density comes from the AIPS task {\it TVSTAT} 
\item Column 9-10: Beam deconvolved angular sizes for resolved sources in 
             arcsec. For multi-component sources the source total size is
             listed. For clearly extended sources  the size was measured 
             directly on the radio images from the lowest reliable contour. 
             Details on the source structure are given in 
             Section~\ref{comments}
\item Column 11: Source major axis position angle, in degrees, measured
             fromNorth through East 
\item Column 12: Notes relative to the radio morphology based on the 
             criteria given at the beginning of Section~\ref{sample}. 
             For sources reliably fitted by a Gaussian model the following 
             letters are used: ~$u~=$ unresolved, ~$ru~=$ partially resolved, 
             ~$r~=$ resolved (see \ref{OPTS}). Multi--component sources or 
             sources too extended for a Gaussian fit are 
             indicated by $''ext''$.
\end{itemize}

%------------------------------------------------------------------
%Fig 2
\begin{figure}
 \centering 
 \includegraphics[bb=0 120 574 574, width=14cm]{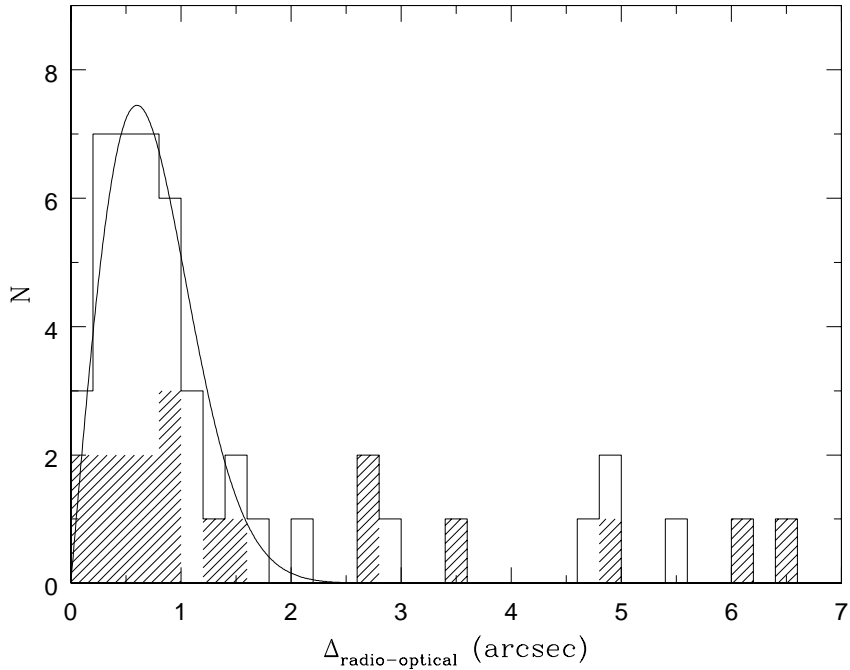}
   \vspace{-1cm}
\caption{Radio-optical positional offset in arcsec.  Shaded bins indicate 
extended sources. The overimposed curve is the expected distribution 
represented by a Rayleigh law with $\sigma=$ 0.6\arcsec (see text).}
\label{fig2}
\end{figure}
%------------------------------------------------------------------

%______________________________________________________________
\subsection{Optical counterparts}
\label{opt-id}

Radio  source positions  were cross-correlated with the SuperCOSMOS/POSS\,II 
(Second Palomar Sky Survey) catalogue within a search box of
10$''$. When more than one optical  object was present within 
the search area the one closest to the radio position was considered 
as the most probable counterpart. All radio source contours were then 
overplotted on the optical red Digitized Sky Survey DSS\,II images  and 
visually inspected. Typical limiting magnitudes reached by the POSS\,II are 
B$_{J}\sim$22.5 and R$\sim$20.8 \citep{rei91}. When no optical counterpart 
was present on the POSS\,II  additional  optical information was 
considered if available.

\smallskip\noindent
The SuperCOSMOS catalogue has a claimed positional accuracy of $\pm$
0.2\arcsec at  B$_{J}\sim$19 and R$\sim$18, rising to $\pm$~0.3\arcsec at 
B$_{J}\sim$22 and R$\sim$21. Radio positional errors in right ascension and 
declination have been estimated according to \cite{con98}. The positional 
uncertainty derived  for the faintest point-like sources (S$_P \sim$ 0.17 mJy 
beam$^{-1}$) is approximately $\pm$0.3\arcsec in each coordinate. Therefore 
the combined  positional error is estimated to be $\sim0.5''$. An  optical 
object is  considered  the counterpart of an unresolved  radio source if the 
radio to  optical distance is less then  $\approx 4\sigma$, 
i.e.~$\lesssim$~2.1$''$. 
The distribution of the radio-optical positional offsets is given  
Fig.~\ref{fig2}.  Shaded areas indicate extended sources (i.e. $ru$, $r$,
and $''ext''$ in Tab. \ref{tab2}).  The curve overimposed to the observed 
distribution in Fig.~\ref{fig2} indicates the expected distribution, 
represented by a Rayleigh law \citep[see e.g.][]{de77}, with a $r.m.s.$ of 
the optical to radio  distance equal to 0.6$''$. The expected 
distribution accounts very well for the observed one within 
$~\approx~2''$. Larger radio--optical  offsets either correspond 
to extended sources or presumably to incorrect identifications. 
There are only 6 extended sources in Fig.~\ref{fig2} with an offset larger 
than 2.1$''$. Four of them were used in the radio luminosity function 
(RLF) derivation, namely NEP\,2870 B with an offset of 2.6$''$, NEP\,2870 A 
at an offset of 2.7$''$, NEP\,2770 A with an offset of 3.5$''$ and NEP\,2420 D
with an offset of 6.5$''$ (see Notes on Individual Sources for NEP\,2420 D, 
Section~\ref{comments}).  A careful examination of the available data makes 
us confident that the four radio sources have been correctly identified.  
The remaining two  sources (NEP\,2950 C with an offset of 4.9$''$ and 
NEP\,2420 F with an offset of 6.1$''$) could be less secure identifications. 
However they are not used in the evaluation of the radio luminosity 
function since they are at a distance  R/R$_{A}$ (where R$_{A}$
is the Abell radius) larger than 0.2 from the cluster center.  Forty of the 
radio sources of Fig.~\ref{fig2} have an optical counterpart on the POSS\,II, 
32 of which within a radius of 1.5$''$. Three additional sources (NEP\,5281 A,
B and C) were identified on a I-band image taken at the UH 2.2m telescope by 
\cite{gio04}, who measured the redshifts listed in Table~\ref{tab3}. For 
a fourth source, NEP\,3130, we used the APM/POSS\,I catalogue since the 
source is close to a very bright star which is less dominant than in the 
POSS\,II  plates. The optical counterparts of these last four objects are all
within 1.5\arcsec of the radio position. In total 44 radio sources have
a possible optical counterpart.  

\noindent
Ten of the proposed optical counterparts are classified  as star-like
objects in the red plates of the SuperCOSMOS/POS\,II catalogue, all the  
others are classified as galaxies.  For 15 sources a spectroscopic 
redshift is available \citep{gio03}.  Fourteen of them are cluster members, 
two of the 14 are classified as star-like in the SuperCOSMOSS/POSS\,II  
catalogue. The 44 optical counterparts are listed in 
Table~\ref{tab3}.  The columns contain the following information:

%----------------------------------------------------------------------
\begin{itemize}
\item Column 1: Cluster name
\item Column 2: Source identification letter
\item Column 3-4: Coordinates of the optical counterpart (J2000), from the 
              on-line SuperCOSMOS/POSS\,II red catalogue. The APM/POSS\,I 
              catalogue was used for NEP\,3130. Exceptions are the coordinates
              for sources A, B and C in NEP\,5281, which were  derived from 
              the I-band image taken at the UH 2.2m telescope 
\item Column 5-6: Offset between the optical and the radio coordinates 
              ($\Delta$\,RA and $\Delta$\,DEC in arcsec) 
\item Column 7-8: POSS apparent red (R) and blue (B$_{J}$) magnitudes
\item Column 9: Optical classification according to SuperCosmos/POSS\,II  
              (APM/POSS\,I catalogue for NEP\,3130): G = galaxy, S = star-like
\item Column 10: Spectroscopically measured redshift. Typical uncertainty  
             is $\le$0.001. The  asterisk after the redshift denotes that 
             the radio source is a cluster member. 
\end{itemize}
%______________________________________________________________

\section{Field Contamination and Cluster Membership}
\label{field}
Since spectroscopic redshifts for the optical counterparts are available 
for only 15 NEP radio sources (14 of which are cluster members 
belonging to 11 different clusters) we have to evaluate the contamination of 
field sources in order to correctly derive the RLF. As described below two 
methods were used, namely a statistical estimate of field sources and a 
best-guess allocation of cluster members on the basis of their absolute 
magnitude and optical classification. 

\subsection{Field Radio Source Counts}
\label{counts} 
To correctly determine, in a statistical sense, how many radio sources are 
field objects and how many radio sources are cluster members we need to have 
a knowledge of the background source density as a function of flux density. 
For this purpose we used the VLA-VIRMOS deep field survey source counts  
\citep{bon03} which have good statistics at our frequency and flux density 
levels. At the same time we performed an analysis of our data in order to 
check their consistency with the literature data. The comparison was also 
aimed at verifying our completeness level down to the  limiting flux density. 

\begin{figure}
 \includegraphics[bb=0 0 574 594, width=9cm]{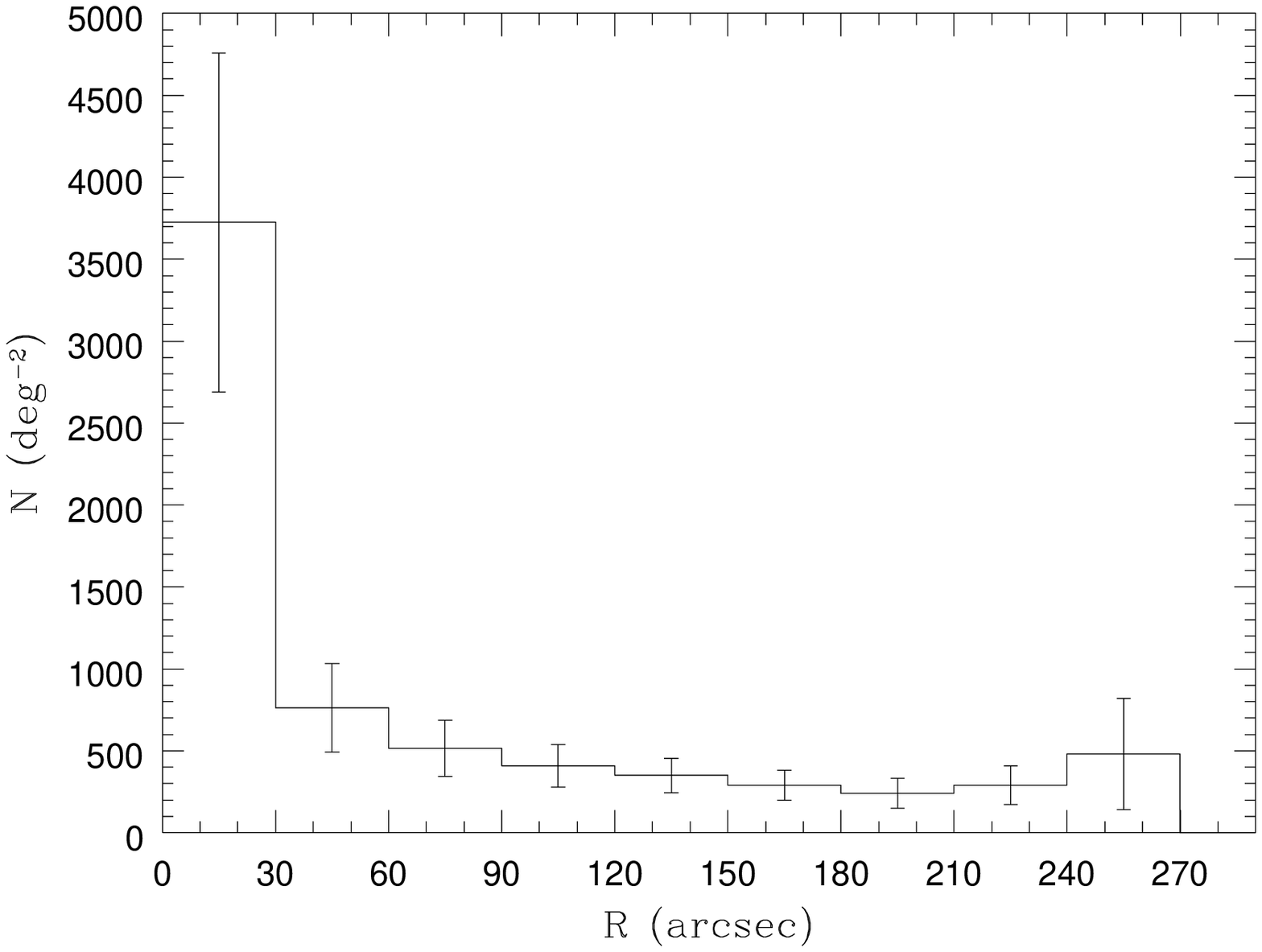}
   \includegraphics[bb=0 0 574 424, width=9cm]{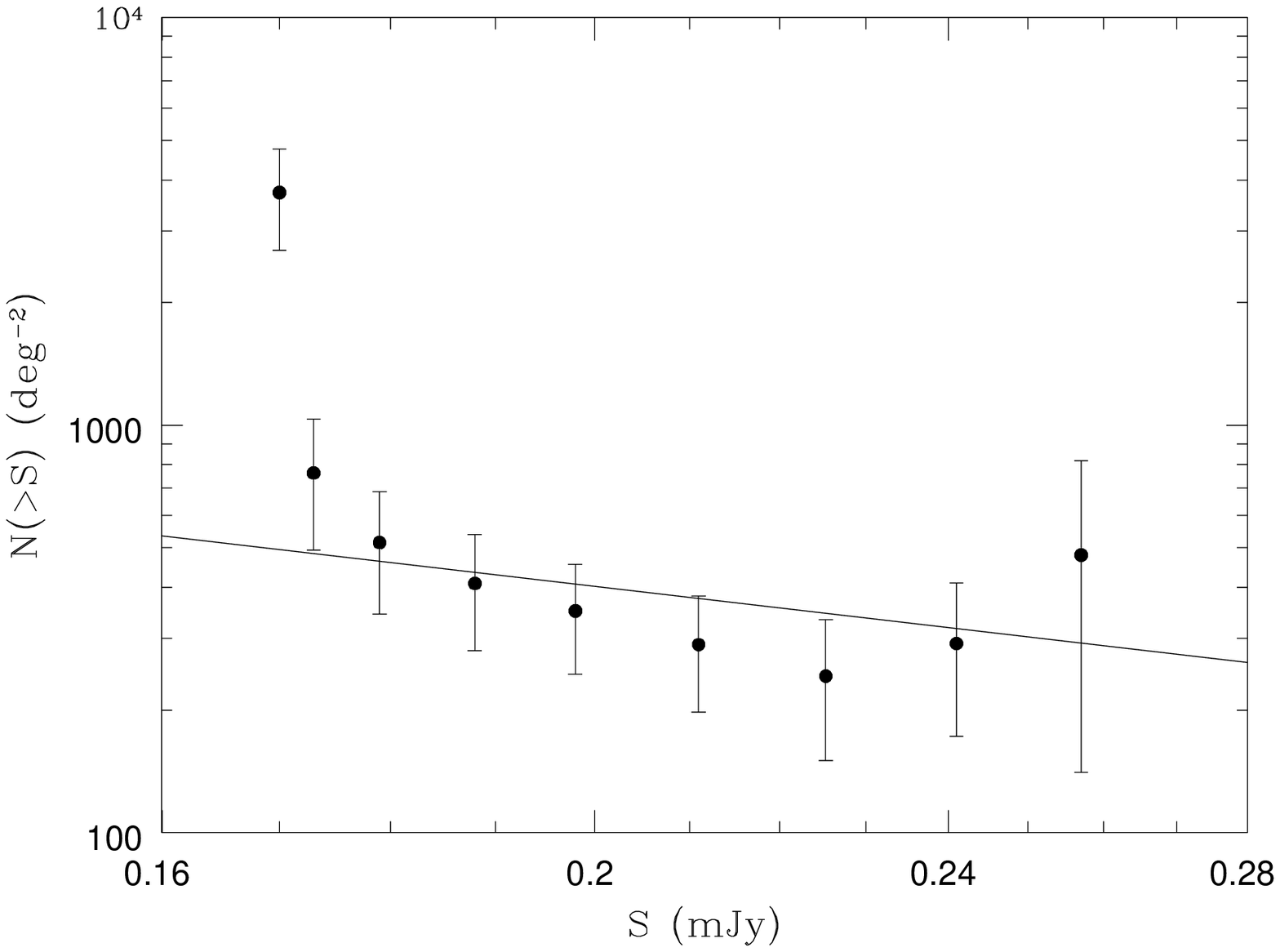}
   \vspace{-1cm}
   \caption{{\em (Left)}: Radial distribution of the radio source density as a 
            function of the projected angular distance from the cluster center.
            {\em (Right)}: Integral radio source counts  of the NEP source 
            sample. The line rapresents the power law fit of the integral 
            logN($>$S)-LogS of the VLA-VIRMOS deep field survey \citep{bon03}.}
   \label{fig3}
\end{figure}

\smallskip
\noindent
We used the complete subsample of 16 clusters with the observed 
brightness limit of 0.17 \mJ and proceeded as follows. The radial 
distribution of the source density at various projected  angular 
distances from the cluster center was obtained by counting objects within 
concentric, $\sim$30\arcsec wide circular annuli centered on the cluster 
itself, and up to a distance of 4.5\arcmin ~from the clusters center. 
The search  radius of 0.5 R$_{A}$ does not extend up to 4.5\arcmin ~for 
all clusters. Thus the more external annuli are not covered by every 
cluster. Therefore the number of sources found in each annulus was normalized 
by the area actually examined. The resulting radial distribution of the 
source density, in the top panel of Fig.~\ref{fig3}, shows a very high
concentration in proximity of the cluster center. 
%Even though field sources 
%are not subtracted, the histogram shows that radio sources in high-z 
%clusters are highly concentrated in proximity of the cluster center 
%similarly to low-z cluster radio sources. The flat distribution at large 
%radii  represents the density of the field sources. 
The data presented in the panel have a bandwidth smearing-dependent 
sensitivity limit folded in. The true flux density limit in each annulus 
decreases as a function of the radial distance according to the curve
shown in Fig.~\ref{figA1} in the Appendix. Therefore  the distribution 
in  Fig.~\ref{fig3} (top panel) converts directly into the integral 
surface density of sources stronger than the flux density limit at the 
middle point of each radial interval shown in the lower panel. Here the 
bandwidth smearing is accounted for by considering the varying flux density 
limit of each annulus. It has to be noted that although these are integral 
source counts all points are independent of each other as they refer to 
different sky areas. Essentially it is like  having many integral 
log N($>$S)-Log S down  to different flux density limits in independent 
regions of sky. We believe that this is a good way  to show the background 
counts in our data, and  compare them with literature  data. 

\smallskip\noindent
Similarly to the radio source density of the top panel, the first two points 
of the NEP log N($>$S)-Log S in the lower panel shows an overdensity  of 
radio sources  due to the presence of the cluster and a flattening as one 
moves to higher flux densities, representative of the field sources. The solid 
line in the bottom panel represents the power law fit of the integral 
logN($>$S)-LogS of the VLA-VIRMOS deep field survey \citep{bon03}.  
Within the errors, the slope of the integral VIRMOS counts (-1.28) and 
its normalization are a good representation of the radio counts of the 
NEP field sources indicating that the NEP  sample is statistically complete.

\begin{figure}
\centering
   \includegraphics[bb=0 0 574 594, width=14cm]{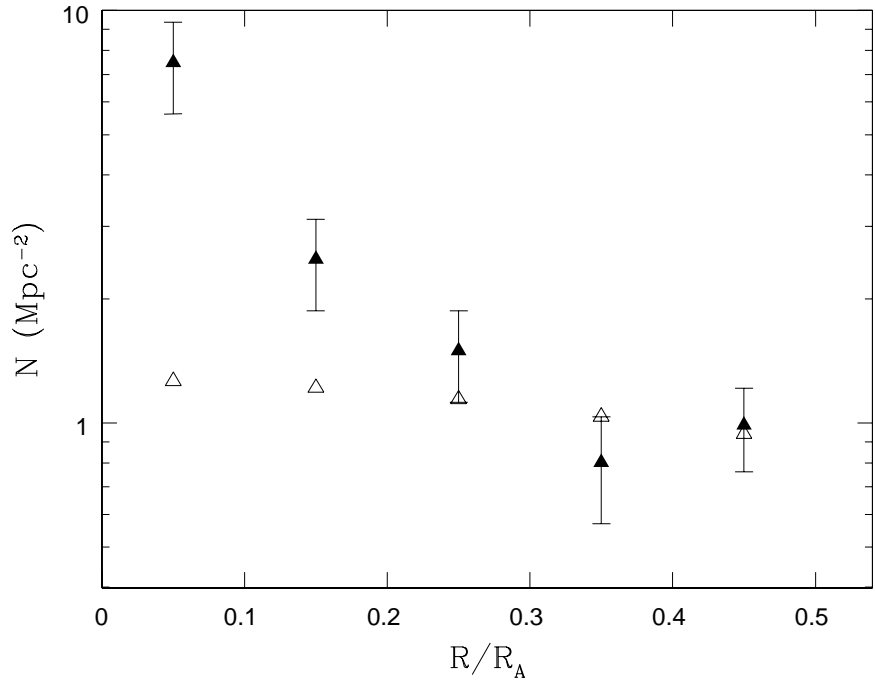}
   \vspace{-4cm}
  \caption{NEP radio sources counts (solid triangles) per Mpc$^{2}$ in 
          bins of R/R$_{A}$ and the expected field contaminations
          (open triangles) estimated as described in the text, 
          Section~\ref{counts}. The errors on the open triangles 
          are of the order of 2\% and are omitted since they are smaller 
          than the symbols.}  
      \label{fig4}
   \end{figure}

\smallskip\noindent
We then proceeded to evaluate the contamination due to non-cluster sources 
in the following way. For each cluster (including NEP\,3130) five radial
intervals in R/R$_{A}$, in steps of 0.1 were defined. Using the background 
source density we computed for each cluster the expected number of sources 
in each annulus above the flux density limit corresponding to the mean radius 
of that annulus. The number of non cluster sources in the various $R/R_A$ 
bins can be obtained by summing in each annulus the expected numbers from
the 17 clusters. The results are an estimate of $\approx$~10
non-cluster sources for $R/R_A \le$ 0.2, and $\approx$ 45 non-cluster 
sources for 0.2$< R/R_A \le$ 0.5, to be compared with the numbers actually 
found of 32 and 47, respectively. Consequently all the sources at 
$R/R_A \ge$ 0.2 are essentially non-cluster sources. In order to stress 
that the source excess is in the inner radii we plot in  Fig.~\ref{fig4} 
the number of NEP radio sources per Mpc$^{2}$ in bins of R/R$_{A}$ (the
solid triangles) and the expected field contaminations (the open triangles).
An examination of the plot shows that the source excess lies in the first 
two bins. For this reason we will limit our analysis only to sources within 
$R/R_A =$ 0.2.
%------------------------------------------------------------------
%Fig 3
%------------------------------------------------------------------   
%------------------------  Fig 4  ------------------------------------------
%Fig. 4
%------------------------------------------------------------------
%------------------------  Fig 5  ------------------------------------------
%Fig. 5
\begin{figure}
\centering
   \includegraphics[bb=0 0 574 594, width=14cm]{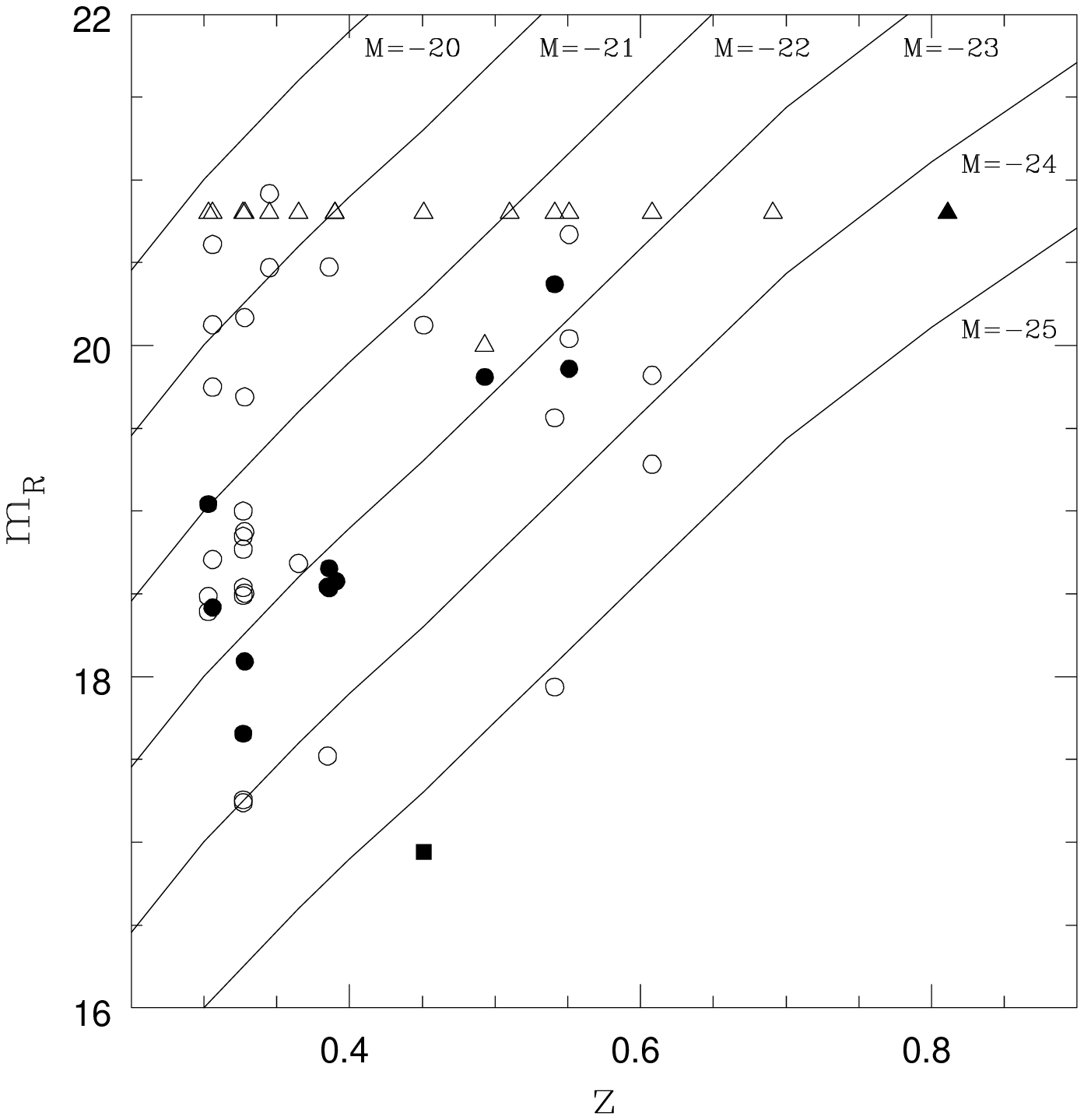} 
   \includegraphics[bb=0 0 574 424, width=14cm]{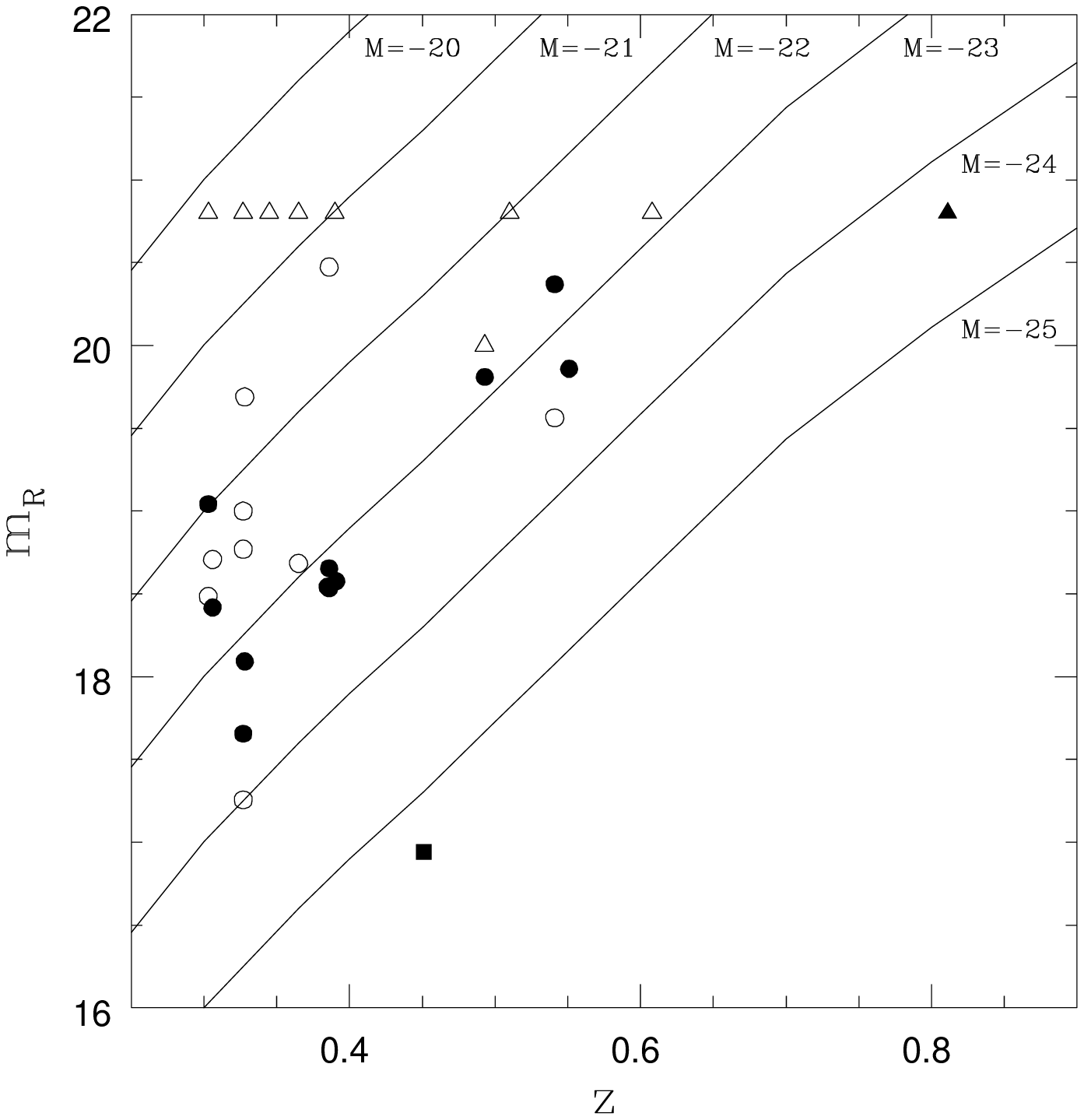}
   \vspace{-4cm}
   \caption{({\em Top}): Apparent red band magnitude of all the NEP radio
   source counterparts as a function of redshift. ({\em Bottom}) panel
   refers only to radio sources within 0.2 R$_A$ from the cluster center.
   Solid lines indicate constant absolute magnitudes. Filled circles and 
   triangles (some are overplotted) indicate cluster members. The filled 
   square indicates a foreground source as explained in the text 
   (Section~\ref{optical}). Open circles indicate 
   radio sources with an optical counterpart in the POSS\,II or POSS\,I 
   survey, open triangles (some are overplotted) indicate radio sources 
   with no optical counterparts, and for which the red  
   magnitude limit of the POSS\,II or POSS\,I has been assumed. }
      \label{fig5}
   \end{figure}
%------------------------------------------------------------------

\subsection{Optical Magnitude Criterion}
\label{optical}

In order to be conservative on the estimate of cluster membership we adopt
a stringent optical criterion  to discriminate between cluster members and
field  sources based on the cluster galaxy absolute magnitude and their
optical  classification.
Assuming that each radio source is at the redshift of the associated
cluster, the  absolute red magnitude of each object, $M_r$, has been
calculated from its apparent magnitude with the {\it K}-correction  
applied as in \cite{fuk95}. The magnitude limit of the POSS\,II, or
POSS\,I surveys has been assigned as upper limit to those radio sources 
with no optical counterpart, except for the radio source NEP\,3130 C
which is  obscured by a bright star in both surveys.

\medskip
\noindent
Figure \ref{fig5} (top) shows the red band apparent magnitude of the sample 
radio sources as a function of redshift.  Solid lines  indicate constant 
absolute magnitudes. Filled circles refer to sources with a spectroscopically 
measured redshift. The filled triangles (some are overplotted)  indicate 
cluster sources with redshift  but with no optical counterpart on the POSS. 
The filled square represents a galaxy towards the cluster NEP\,4560
(z$=$0.451) which has a measured spectroscopic redshift of 0.174, but is
plotted here as though it were at the redshift of the cluster. Open  circles 
indicate radio sources with an optical counterpart in the POSS\,II or POSS\,I
survey, open triangles (some are overplotted) indicate lower limits to the 
magnitude. Bottom panel is the same as top panel but with only  the sources 
within  R/R$_A<$ 0.2.  Analysis of this plot suggests that all bona-fide 
cluster members lie in the range $-25 < M_r < -$21.  The only non-cluster 
member with a measured redshift (the filled square) lies outside this 
range. 
The bottom panel plot is a convenient tool to estimate from the apparent
magnitude which of the radio sources without redshift could fall in the  
$-25 < M_r < -$21 range and thus be considered cluster members. It is also 
clear from the  figure that there are more contaminations at low z, as 
expected, since the area occupied by the low-z clusters is larger.  
%---------------------------Fig  6  ---------------------------------------
%Fig. 6
\begin{figure}
\centering
\includegraphics[bb=0 0 574 574, width=14cm]{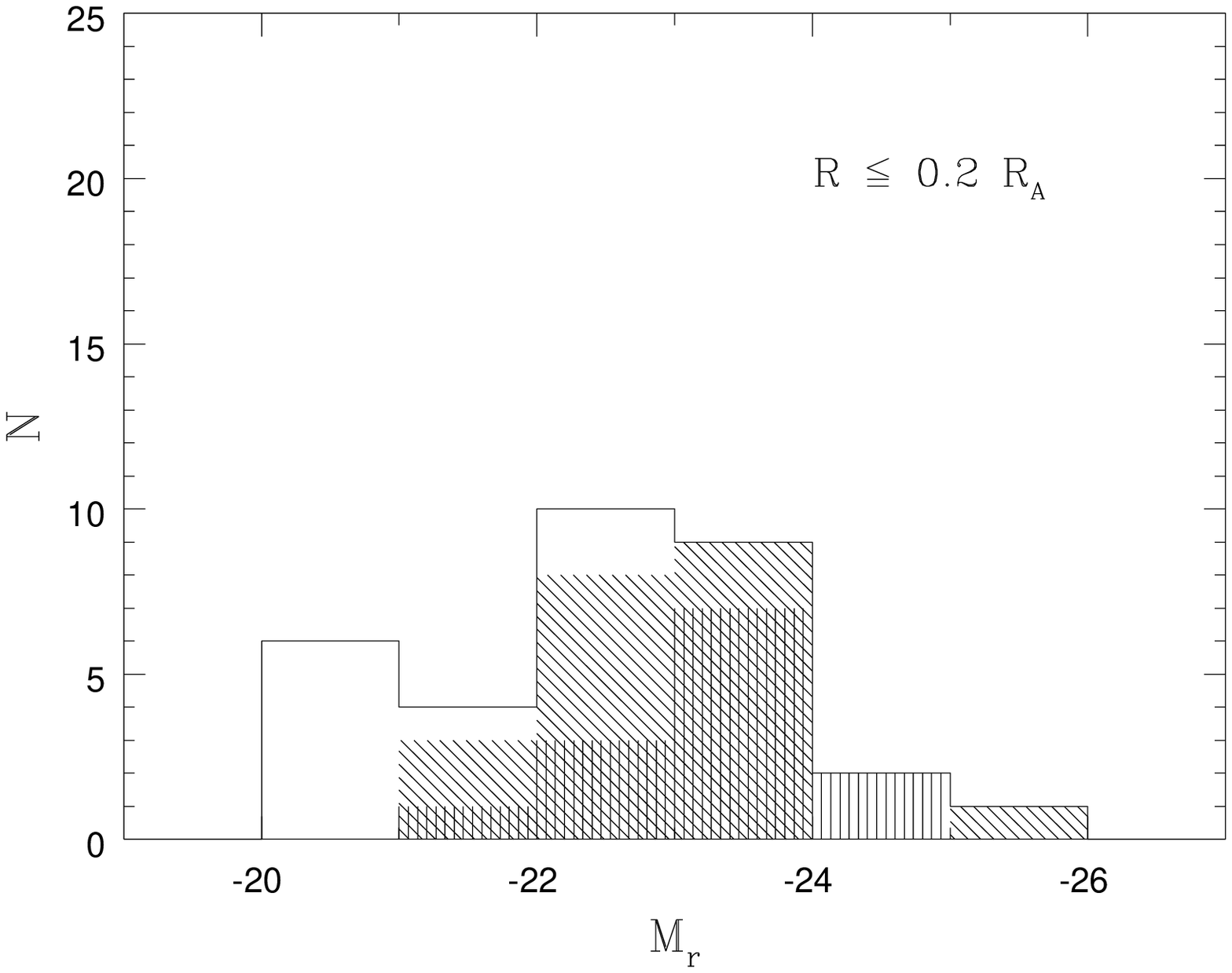}
\includegraphics[bb=0 100 574 394, width=14cm]{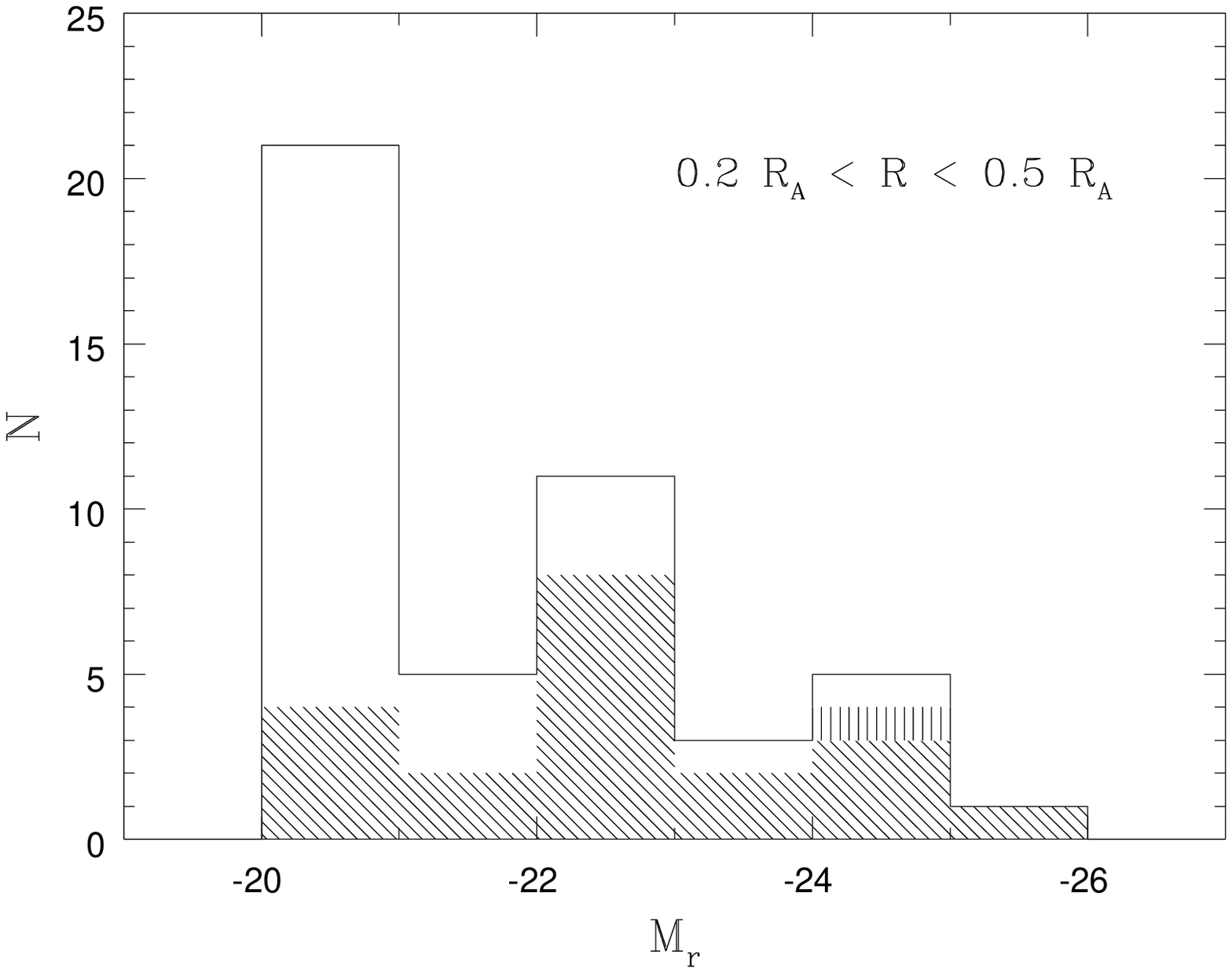}
\vspace{-1.5cm}
\caption{Distribution of the absolute red magnitudes assuming all the 
    radio sources are cluster members. {\bb Histograms shaded by diagonal 
    lines represent sources with a POSS\,II-red 
    or POSS\,I-red optical counterpart. Areas shaded by vertical lines
    indicate all the bona-fide cluster members. Unshaded areas represent radio 
    sources with no optical counterpart. In this case the limiting magnitude 
    of the POSS survey was taken as lower limit apparent magnitude for the 
    objects (including the three galaxies of NEP\,5281 A, B in the top panel 
    and  NEP\,5281 C in the bottom panel at $-25 < M_r < -24$ for which a 
    redshift is available).} {\em (Top)} panel refers to the 32 
    radio sources within 0.2 R$_A$ from the cluster center. 
    The object at $-26 < M_r < -25$ refers to a galaxy towards the cluster 
    NEP\,4560 (z$=$0.451) which has a measured spectroscopic redshift of 
    0.174, but its M$_{r}$ was computed as though it were at redshift of the 
    cluster; {\em (Bottom)} panel refers to 46 radio sources at a distance 
    larger than 0.2 R$_A$. One source, NEP\,3130 C, is missing since it is 
    obscured by a bright star in both  POSS\,I and POSS\,II plates.}
   \label{fig6}
   \end{figure}
%------------------------------------------------------------------
The histograms in Fig.~\ref{fig6} show the distribution of the absolute 
magnitude for all the radio sources with R/R$_{A} \le 0.2$ (top histogram) 
and R/R$_{A}>0.2$ (bottom histogram). Histograms highlighted by diagonal 
lines in both panels indicate radio sources with an optical counterpart in the 
POSS\,II-red or in the POSS\,I-red plates. {\bb Areas highlighted by
vertical lines indicate all the bona-fide cluster members with a spectroscopic 
redshift. The three objects at $-25 < M_r < -24$  with vertical lines 
are members of the z $=$ 0.81 cluster NEP\,5281 and do not have a 
POSS\,II counterpart but are visible in the ESO DSS\,II infrared images 
and on the I-band UH 2.2m telescope image. Thus only lower limits to the 
red and blue magnitudes are available. The object in the top panel 
at $-26 < M_r < -25$, indicated by the filled square in Figure \ref{fig5}, 
is the foreground galaxy at z$=$0.174. The top panel clearly shows that  
the thirteen bona-fide cluster members (vertical lines) are in the range  
$-25 < M_r < -21$.}  According to previous results 
\citep[e.g.][]{lo96} it is likely that the remaining genuine cluster sources 
are in this  same  magnitude range and that objects with $M_r \ge -21$ are 
mostly background  objects.  Finally, three sources within $R/R_A \le 0.2$
satisfying the optical magnitude criterion are classified as star-like 
objects and are possible non-cluster galaxies. However, we cannot exclude 
that they are genuine cluster members or background quasars. It follows that 
up to 10 sources within $R/R_A = $ 0.2 are not cluster members.  
This estimate is consistent with the estimate of $\approx$ 10 obtained  in 
Section~\ref{counts}. On the other hand, the bottom histogram of 
Fig.~\ref{fig6} ($R/R_A  > 0.2$), when allowance is made for the upper 
limits, does not show  any preferred magnitude, therefore the majority of 
these sources  are most probably non-cluster sources, in agreement with the 
estimate of Section~\ref{counts}. In conclusion we estimate a number
of 22 genuine cluster radio galaxies.

\section{Radio Morphology}
\label{morph}

\begin{figure}
\centering
   \includegraphics[bb=0 120 574 580, width=14.5cm]{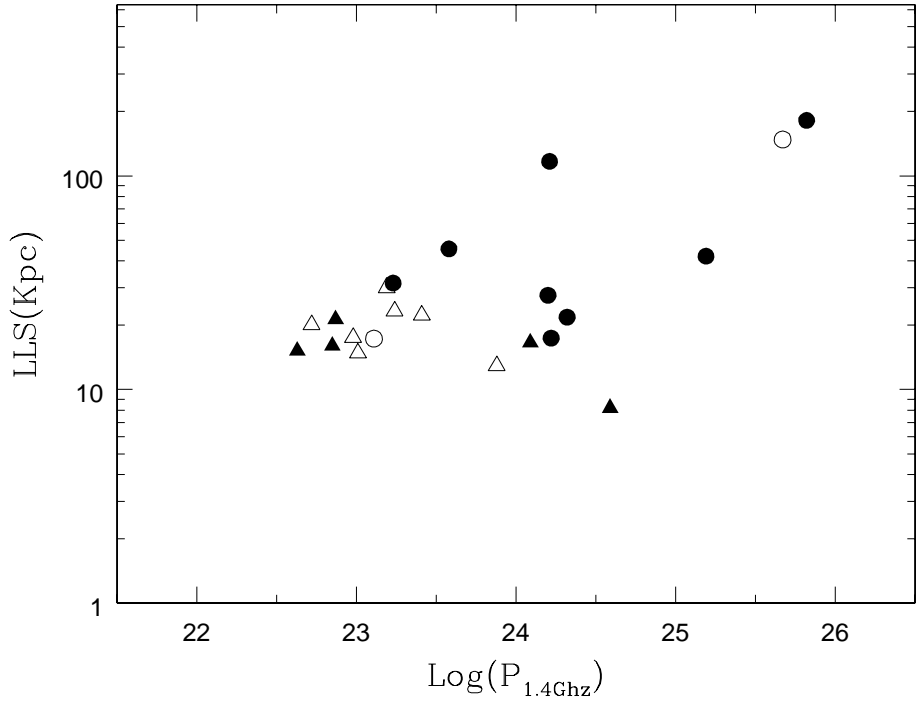}
   \vspace{-0.2cm}
   \caption{Largest linear size in kpc versus Log P$_{1.4Ghz}$ for 
      cluster members with redshift (filled symbols) and sources
      considered cluster members according to the optical criteria
      (open symbols). Circles refer to resolved sources  and triangles 
      to unresolved sources with upper limit to the size (see 
      Section~\ref{morph}).
      }
  \label{fig7}
   \end{figure}

Approximately  68\% of the NEP radio sources are unresolved with an estimated
maximun  deconvolved angular size of $\sim$1.2$\times$beam size for fainter 
sources ($\sim$ 0.2 mJy), and $\sim$0.33 $\times$beam size for sources with a 
flux density $\ge$ 1 mJy. Approximately 16\% (13 sources) have angular 
sizes in the range $\sim 8\arcsec$ to $\sim 40$\arcsec and the remaining 
sources have sizes less than 8$''$.  Five of the 14 sources labelled 
as  $``ext''$ in  Table~\ref{tab2} exhibit a head--tail morphology, four are
double sources, one is a triple source and four have an uncertain 
morphological classification.  Fifty percent of the 14 $``ext''$ sources are 
within 0.2 R$_A$  and five of them are associated with 
the brightest central galaxies, or BCGs. Four more sources (one classified 
as $``r''$ and three classified as $``u''$) coincide with the BCG. In total 9
radio sources are associated with the cluster brightest central  galaxies, 
that is about half of the 17 clusters have a radio emitting central galaxy. 
If we limit our analysis to the assumed 22 cluster radio galaxies the resolved
and  unresolved sources divide equally. Although the statistics are very small 
there is a correlation between source linear size and radio power as 
shown in Fig.~\ref{fig7}. Filled symbols indicate cluster radio sources with 
redshift, open symbols indicate radio sources considered to be cluster 
members according to the optical magnitude criterion. The circles indicate 
the resolved, partially resolved and extended sources while the triangles 
refer to unresolved sources. The sizes of the unresolved sources are obtained 
by assigning appropriate upper limits to the maximum deconvolved angular 
sizes (see Section~\ref{sample}). About half of the sources with 
log P$_{1.4}\ge$  24.0 has a linear size  $LLS>$ 40 kpc, while most of the 
sources with  log P$_{1.4} \le$ 24.0 have a linear size less than $\sim$ 30 
kpc. Fig.~\ref{fig11}, ~\ref{fig12} and ~\ref{fig13} show radio contours of 
the more extended radio sources overplotted onto optical  POSS\,II red images 
(greyscale). For NEP\,5281 C the I-band image obtained at the UH 2.2m 
telescope was used. In all images North is up and East to the left. 
%----------------------- Fig 7 ------------------------------------------
%Fig. 7
%------------------------------------------------------------------

\subsection{Notes on Individual Sources}
\label{comments}

-- NEP\,1730 B: This source has a component with S$_I = $ 0.56
        mJy, which coincides with the optical BCG, plus a second fainter
        component with a  S$_I = $ 0.22 mJy. 

\smallskip\noindent
-- NEP\,1730 D: Small double source with no optical counterpart.
The radio source component flux densities are S$_I = $ 0.19 ~mJy (North)
and S$_I = $ 0.17 mJy (South). 

\smallskip\noindent
-- NEP\,1730 F: This radio source is associated with a weak galaxy 
(m$_{R}=$20.6) coinciding  with the source peak which was considered 
the identification (see top panel of Fig.~\ref{fig11}). The radio 
morphology resembles a  small head-tail  source thus not excluding 
its cluster membership in spite of  the faint galaxy  red magnitude 
(Section~\ref{optical}). There is a similar magnitude object 
coinciding with the South-West radio extension of the source.

\smallskip\noindent
-- NEP\,2420 A: Located at the  center of the cluster A2280, this is a triple
straight source with a linear size of 116.8 kpc (the southermost source in 
bottom panel of Fig.~\ref{fig11}). Although  we have no  radio spectral 
information we tentatively assume that the central component is the radio 
core. The lobes appear oriented roughly perpendicular to the source major 
axis. A higher resolution map would be required to properly describe the 
source structure, in particular  to check if the brightness peaks are 
hotspots. In this case the source could be classified as an FRII, quite 
unusual for a central cluster radio source. The radio source components 
flux densities are:  S$_I = $ 2.72 mJy (East), S$_I = $ 0.31 mJy (Center) 
and S$_I = $ 3.39 mJy (West). 

\smallskip\noindent
-- NEP\,2420 D: The overall head-tail morphology of the source is 
dominated by two bright components, a very extended emission to the East 
of the two main components and a slight extension to the West. At first
sight an inspection of Fig.~\ref{fig11} could give the impression that 
NEP\,2420 D and the source to the East NEP\,2420 E described below are 
components of the same single radio source hosted by the bright object at
17$^{h}$43$^{m}$31.5$^{s}$ $+$63$^{\circ}$42$'$51.9$''$.
However the object is classified as a star by SuperCosmos and APM surveys
and it also shows spikes (confirming its stellar nature) in much deeper 
proprietary images taken with the Canada-France-Hawai$'$i Telescope (CFHT) 
that we can access. In addition both NEP\,2420 D and E have an head-tail 
morphology which goes against the double radio source hypothesis.
\cite{la93} studied the field of  NEP\,2420 D and identified this radio source
with a faint galaxy about 4$''$ North of the midpoint of the radio hotspots. 
In \cite{la99} the radio source is identified  with a faint object at the 
midpoint of the radio hotspots. Both radio galaxies are consistent, according 
to Lacy and collaborators, with being members of the A2280 cluster at 
z$=$0.324. However, \cite{la99} are surprised by the fact that, whichever is 
the identification, the radio source is very unusual in having an apparently 
very subluminous optical counterpart. This consideration prompted us to 
examine the whole A2280 area by inspecting CFHT images much deeper 
than the DSS\,II. We believe that the correct identification is the brighter 
galaxy to the West which has an {\bb absolute} magnitude comparable to the 
other cluster  galaxies (see Table~\ref{tab3}). Please note that the radio 
contours  are clearly distorted and denser at the location of this galaxy in 
our  Fig.~\ref{fig11}.
\cite{la92} could barely detect either the very extended tail to 
the East and the extension to the West given their A-array high-resolution 
and low sensitivity observations (see Figure 5 of their paper  where 
only the structure of this double radio source, called 8C\,1743$+$637, 
is clearly outlined).  In conclusion NEP\,2420 D is classified by us as 
a head-tail radio source and the optical galaxy to the West is assumed to 
be the  source optical  counterpart. The radio source has  been used in the 
evaluation of the RLF since it lies  within 0.2 R$_{A}$ from the cluster 
center. {\bb We have also considered the case where the bright object in 
between the two radio sources D and E in Fig.~\ref{fig11} is a field quasar 
and thus is the optical identification. The RLF remains unchanged within 
the errors.}

\smallskip\noindent
-- NEP\,2420 E: The radio source appears to have a head-tail morphology
(eastern source in bottom panel of Fig.~\ref{fig11}) and it is very likely 
associated to the cluster given the absolute optical magnitde of the 
associated galaxy. Since the source is located at $R/R_A$ = 0.21, 
it was not used for the computation of the radio luminosity function. 

\smallskip\noindent
-- NEP\,2420 F: The source is elongated in North-South direction and 
does not show a well defined morphology. The galaxy at  $\sim$6\arcsec to 
North-East of  the radio peak is suggested as the radio source optical 
counterpart. On the basis of the magnitude criterion of Section~\ref{optical}
this galaxy might be a cluster galaxy.

\smallskip\noindent
-- NEP\,2420 H: The radio source shows an amorphous morphology without any 
visible optical counterpart.

\smallskip\noindent
-- NEP\,2770 A: The radio source is located at the cluster center and  
shows a bright peak and a faint tail to the South. The galaxy identified 
as the optical counterpart is a cluster member and is located 
$\sim$3\arcsec to North-East of the brightness peak. The source could  
be classified as a small head-tail source but a higher resolution map is 
required to confirm this interpretation.

\smallskip\noindent
-- NEP\,2870 A: This is a double radio source at the cluster center
with faint extensions off axis to the North (see top panel of 
Fig.~\ref{fig12}). The object suggested as the optical counterpart is 
classified as star-like in the SuperCOSMOS/POSS\,II catalogue. For this 
reason we have not used it in the computation of the RLF (Section~\ref{RLF}).
Should it be a compact galaxy according to the magnitude criterion of 
Section~\ref{optical} then it might be a cluster member.
	    
\smallskip\noindent
-- NEP\,2950 C: Double radio source whose component flux densities
are: S$_I =$ 0.35 mJy (North) and S$_I =$ 0.44 mJy (South).
The assumed optical counterpart, located close to the peak of the brighter 
southern component, is classified as a star-like object in the 
SuperCOSMOS/POSS\,II catalogue. 

\smallskip\noindent
-- NEP\,3130 A: The optical image is shown in the bottom panel of 
Fig.~\ref{fig12}. The radio source is located at the cluster center.

\smallskip\noindent
-- NEP\,3200 A: This object is at the redshift of the cluster and
shows a broad H$\alpha$ emission line in its optical spectrum. It is 
thus classified as a broad line radio galaxy. According to the SuperCOSMOS
catalogue the galaxy is blue with a B$-$R$=$0.8.

\smallskip\noindent
-- NEP\,3450 A: This is the largest linear size radio source of the sample
($LLS\sim$180 kpc; see Fig.~\ref{fig13}). It is at the cluster center and  
shows a double radio structure of the FRI type. 

\smallskip\noindent
-- NEP\,4150 H: This source is a second example of a broad emitting line
object at the cluster redshift. The optical spectrum shows  Balmer lines 
with FWHM $=$ 2500 km s$^{-1}$ and it is thus classified as a 
broad line radio galaxy. According to the SuperCOSMOS catalogue the 
galaxy  is blue with a B$-$R$=$1.3.

\smallskip\noindent
-- NEP\,5281 C: This radio source was identified on a I-band image
of the UH 2.2m telescope taken specifically for this cluster which is the 
most distant cluster in the survey (z$=$0.81). An image is shown in the
bottom panel of  Fig.~\ref{fig13}. The radio source  resembles a small 
head-tail source.

%-----------------------------------------------------------------------
\section{The NEP Radio Luminosity Function}
\label{RLF}

The radio luminosity function  is a powerful statistical 
tool to investigate the radio properties of a galaxy population.
It is reasonable to expect that any difference in the radio properties 
of a sample of galaxies should be reflected in the radio luminosity 
function, either as a change in  shape, amplitude or both.
In absence of optical counts of ellipticals for the clusters under
study it is not possible to determine the fractional 
radio luminosity function (i.e. the fraction of cluster galaxies 
able to produce a radio source) as was done by \cite{lo96}. Following
\cite{fan84} we have adopted the following approach: the RLF represents 
the {\it number of radio galaxies per cluster} as a function of radio power. 
Given the very high contamination by field sources for $R/R_A >$ 0.2 
(see Section~\ref{field}) only  radio sources 
lying inside this radius have been used to derive the RLF. The sample of 32
objects used for the computation of the RLF is listed in Table~\ref{tab4} 
where the columns contain the following information:

%------------------------------------------------------------------
\begin{itemize}
\item Column 1: Internal NEP cluster identification number
\item Column 2: Source identification letter
\item Column 3, 4: {\it K}-corrected  peak and integral radio
      luminosities (see text).  For the {\it K}-correction a spectral 
      index $\alpha = -$ 0.8 (S$_{\nu}$ $\propto \nu^{\alpha}$) has been 
      assumed
\item Column 5: Absolute red magnitude, $M_r$ (see Section~\ref{optical})
\item Column 6: Largest linear size in kpc
\item Column 7: Source weight used for the computation of the RLF (see text)
\item Column 8: Comment on cluster membership, see footnote to the table.

\end{itemize}
%-------------------------------------------------------------------

\begin{figure}
\centering
   \includegraphics[bb=0 0 574 574, width=14cm, height=14cm]{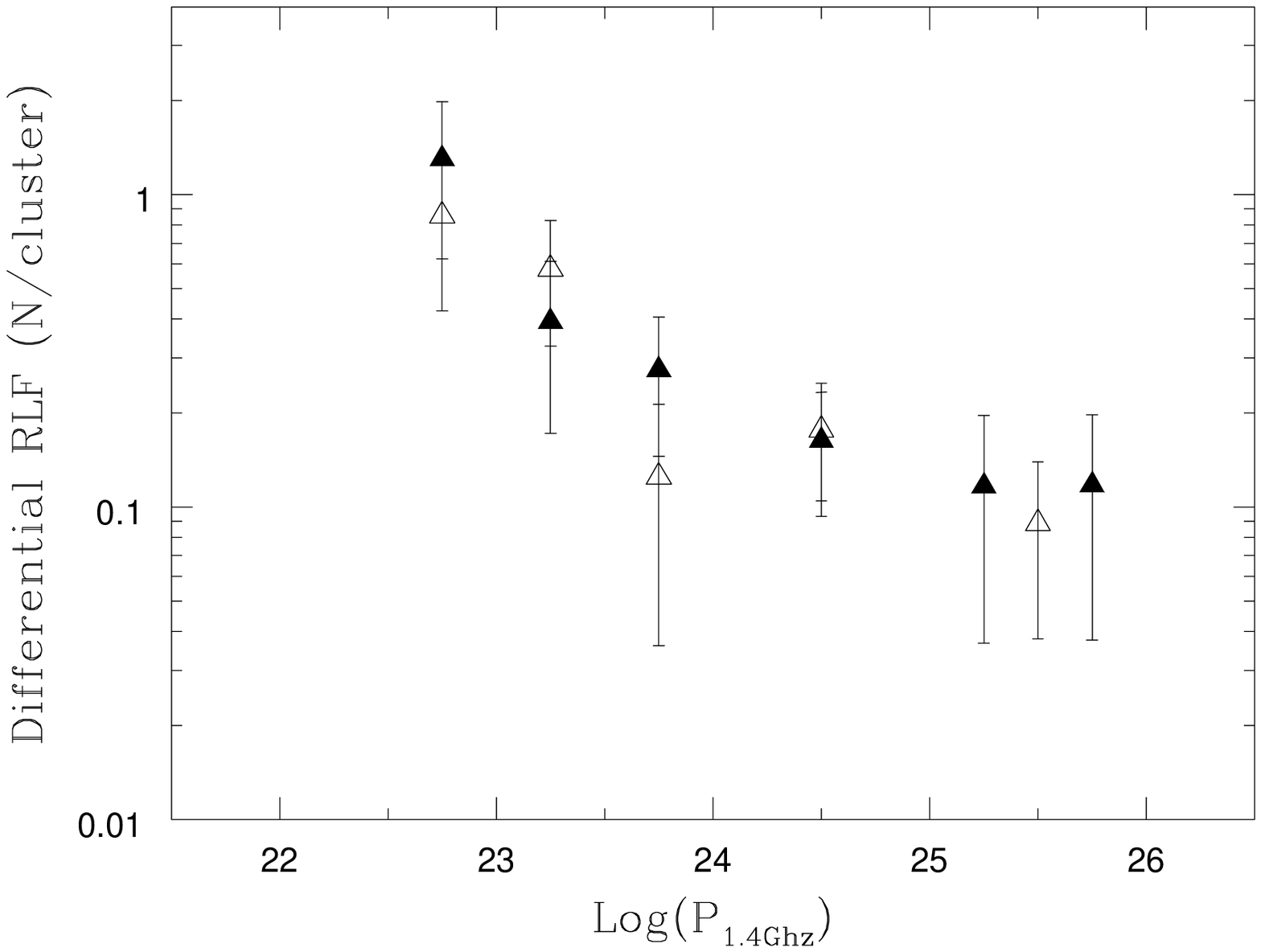}
   \includegraphics[bb=0 0 574 424, width=14cm, height=10.3cm]{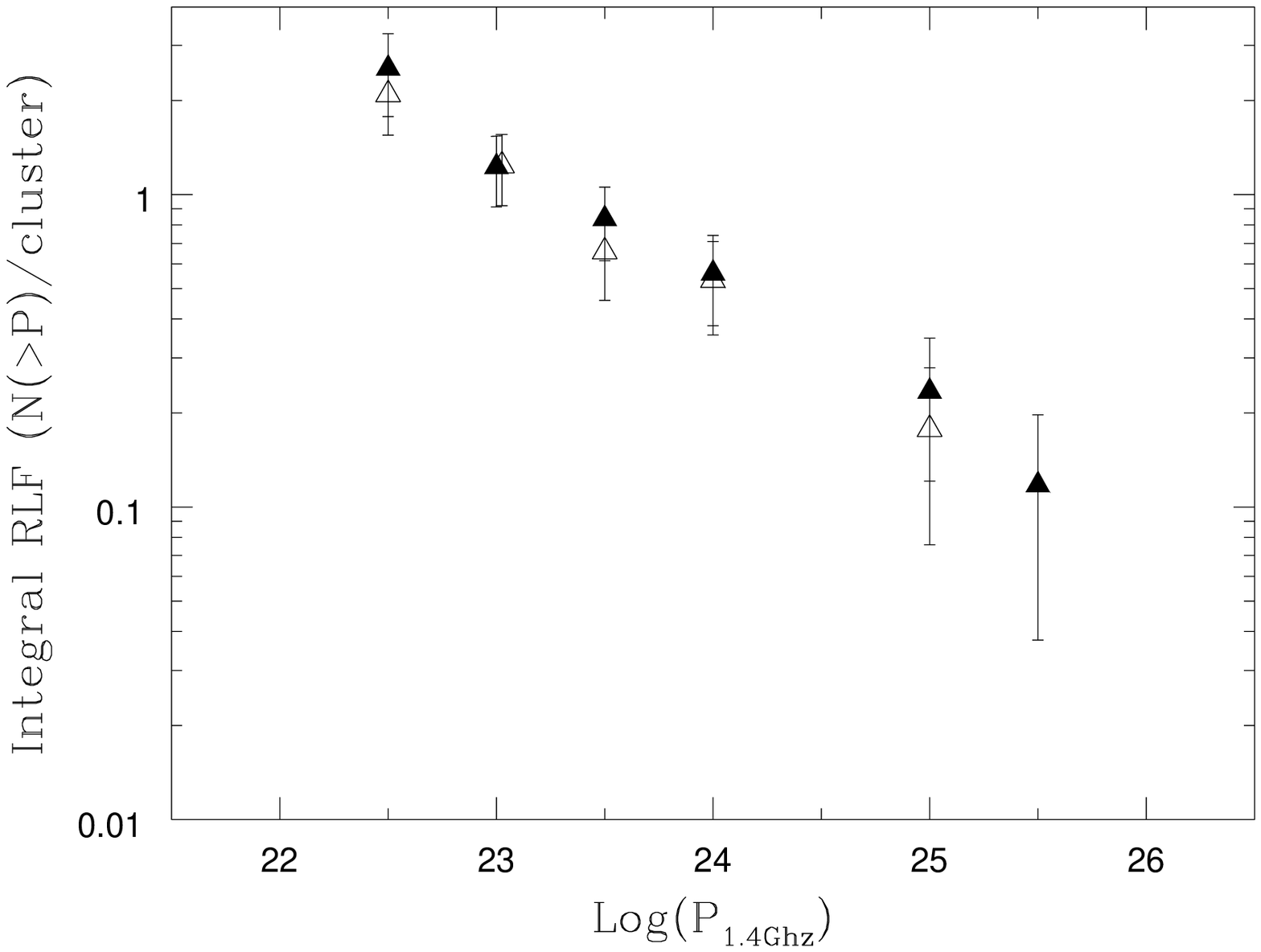}
   \vspace{-4cm}
   \caption{{\em (Top)}: The differential NEP radio luminosity function. 
       {\em (Bottom)}:  The integral NEP radio luminosity function. Open
       triangles represent the RLF computed subtracting the ten,
       either foreground or possible contaminant objects 
       (see Section~\ref{optical}). Filled triangles indicate the RLF 
       computed subtracting the contaminant sources according to the 
       statistical method described in Section~\ref{RLF}. The open triangle 
       in the bottom panel at log P $=$ 23 has been slightly shifted to the 
       right for clarity.  The bins have different widths since they
       were chosen so as to always have at least two objects per bin. Each 
       bin occupation has been normalized to the interval 
       $\Delta$\,log P $=$ 0.5.}
   \label{fig8}
   \end{figure}

\noindent
For each radio source both peak and integral radio luminosities  (P$_{peak}$, 
P$_I$) have been computed.  {\it K}-corrections were  calculated assuming a  
spectral index   $\alpha = -$ 0.8 (S$_{\nu}$ $\propto \nu^{\alpha}$)  and then 
applying the relation {\it K}(z)$=$(1+z)$^{-(1+\alpha)}$.  In the same way a 
luminosity limit $P_{cl,lim}$, corresponding to the  peak brightness
limit of 0.17 mJy beam$^{-1}$, has been calculated for each cluster except
for NEP\,3130, for which its own flux density limit has been used.
The \label{weight} contribution $W_{RLF,i}$ (column 5 of Table~\ref{tab4}) 
of each source {\it i} to the RLF is computed as:
$$W_{RLF,i} = 1/N{cl,i}$$
\noindent 
where $N_{cl,i}$ is the number of clusters in which that source could have
been detected ($P_{peak,i} \ge P_{cl,lim}$). Then the individual source 
contributions within logarithmic bins of  total  luminosity
have been summed.  The bins may have different widths since they
were chosen so as to always have at least two  objects per bin.  Each bin 
occupation has been normalized to the interval $\Delta$\,log P $=$ 0.5. In 
order to  estimate the effect of the $\sim$10 non cluster radio sources  on 
the RLF  two different approaches were adopted. The first one is simply to 
omit sources which are unlikely to be cluster members, namely those 
objects with  $M_r \ge -21$ or $M_r  \le -25$ (seven objects) 
and the three objects which are classified as star-like objects. The second 
approach  is based on the statistical estimate of the field sources 
(see Section~\ref{counts}). The number of contaminant field 
sources,  computed for each cluster as described in Section~\ref{counts}, 
was distributed  into logarithmic bins of flux density according to the 
logN-LogS of the field  sources. The  geometric mean of each flux density 
interval has been assigned to these contaminant "fractions" of sources.  
Their spurious contribution  was computed in the same way as for the radio 
sources and subtracted from  the RLF.  The differential logarithmic RLFs, 
computed with the two methods, are given in Table~\ref{tab5} and 
Table~\ref{tab6}  and are shown in the top panel of Fig.~\ref{fig8}.  
They agree pretty well within the statistical errors. The bottom panel  
of Fig.~\ref{fig8} shows the integral RLFs, also given in Table~\ref{tab5}
and Table~\ref{tab6}.

%------------------------------  Fig 8 ------------------------------------
%Fig. 8
%------------------------------------------------------------------

%---------------------------  Fig  9  ---------------------------------------
%Fig. 9
\begin{figure}
\centering
   \includegraphics[bb=0 0 574 574, width=14cm, height=14cm]{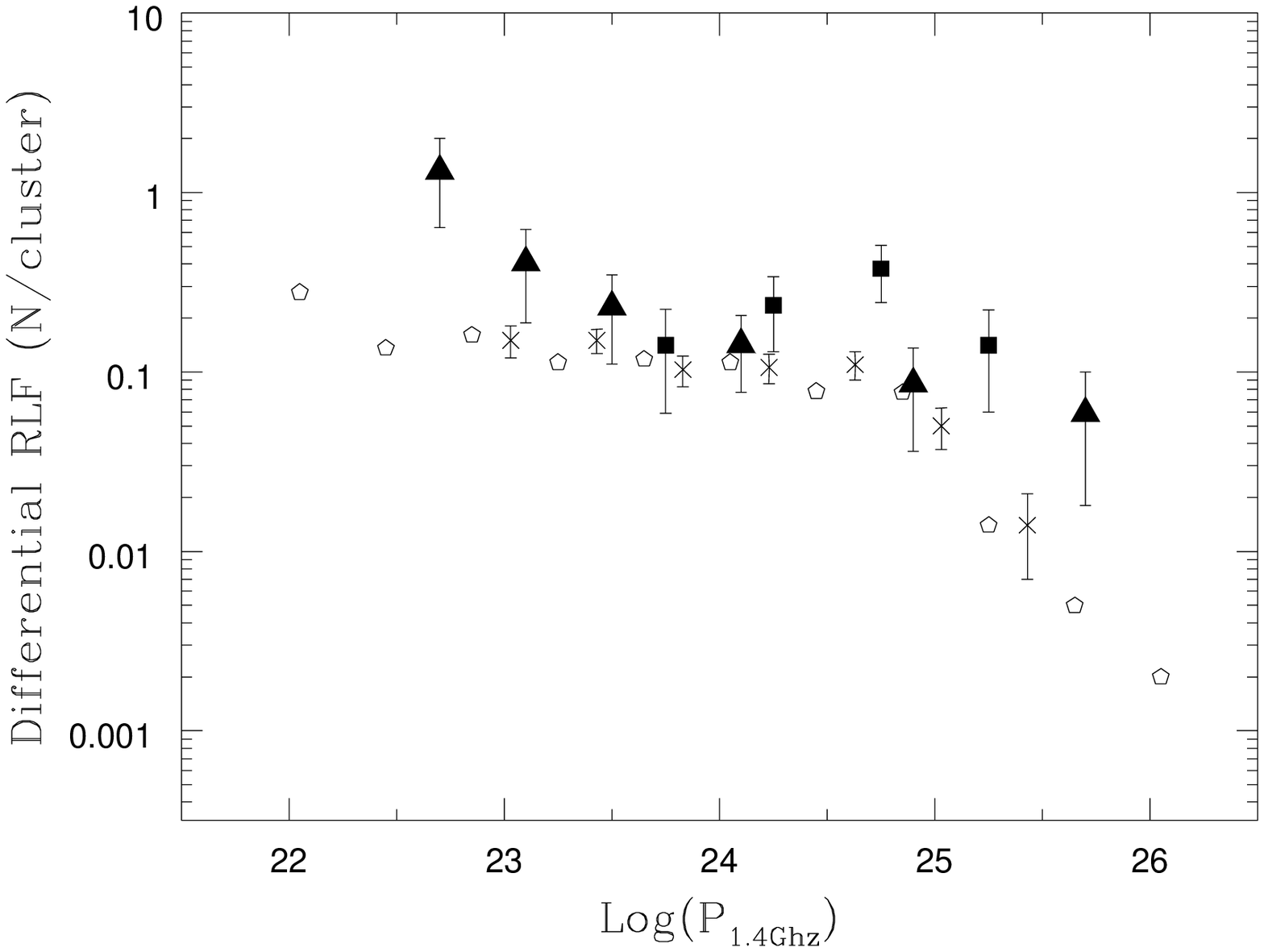}
   \includegraphics[bb=0 0 574 424, width=14cm, height=10.3cm]{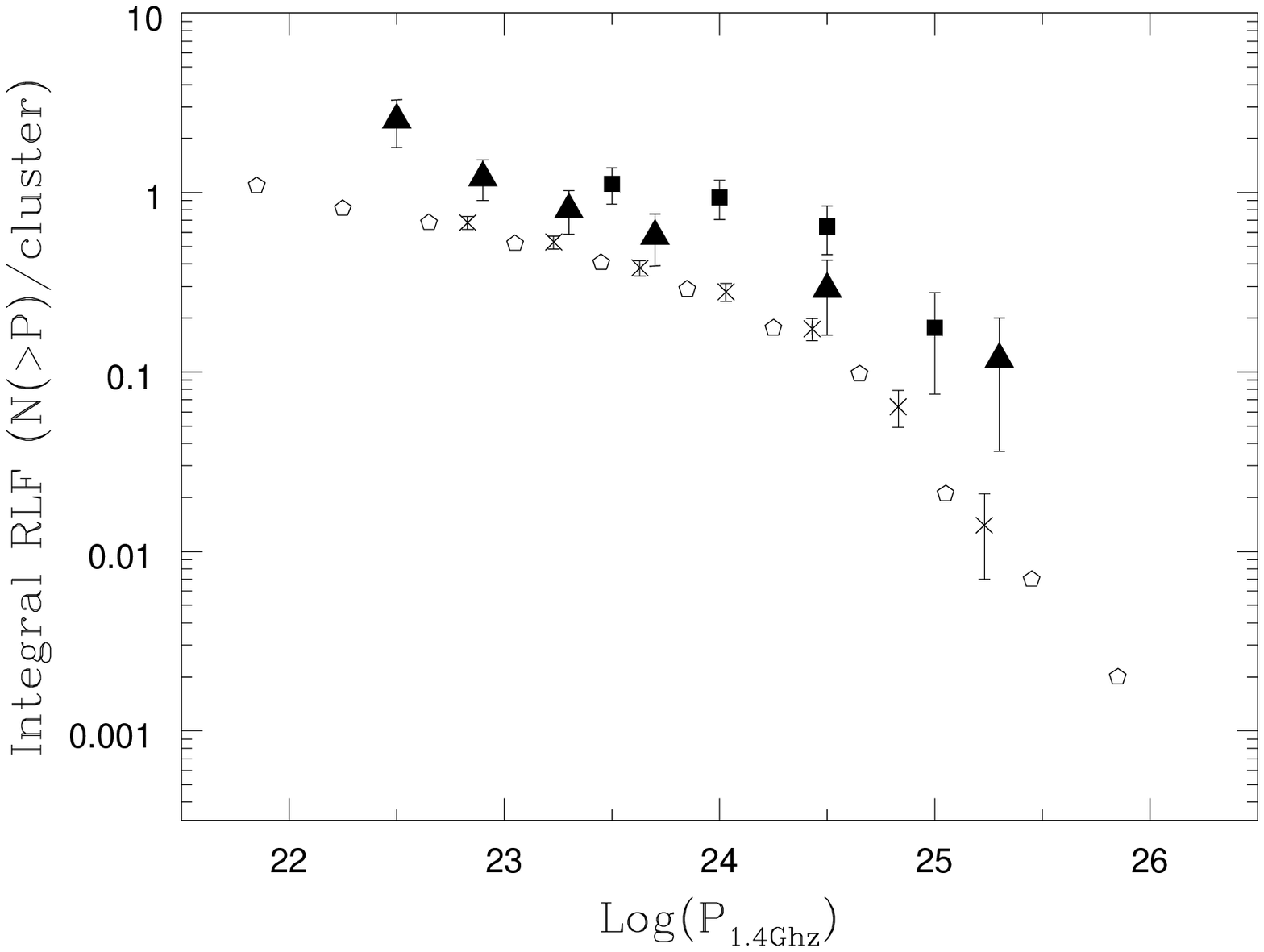}
   \vspace{-4cm}
   \caption{Differential {\it (Top)}  and integral {\it (Bottom)} radio 
   luminosity functions. The distant cluster radio luminosity 
   functions are indicated by filled triangles for the NEP RLF and 
   filled squares for  the RLF derived by \cite{sto99}. The nearby cluster 
   radio luminosity functions are indicated by open pentagons for the RLF 
   derived by \cite{fan84} and crosses for the RLF derived by \cite{lo96}.
   Note that the comparison RLFs refer to R/R$_{A} \le$ 0.3  while the NEP 
   RLF is for R/R$_{A} \le$ 0.2. The bins have different widths since they
   were chosen so as to always have at least two objects per bin.  Each bin 
   occupation has been normalized to the interval $\Delta$\,log P $=$ 0.4.} 
   \label{fig9}
   \end{figure}
%------------------------------------------------------------------

\section{Comparison with Other Cluster RLFs}
\label{RLF-comp}

To verify if any similarities exist in the radio  properties  of local and  
distant clusters of galaxies, the RLF of the radio galaxies of the  NEP 
cluster sample has been  compared with the radio luminosity function  
of distant clusters by \cite{sto99} (see column 2 of their Table 8) 
reanalized using our method. We have also compared the NEP RLF 
with the  RLFs for nearby rich Abell clusters. For this purpose we used 
the RLF by \cite{fan84} and \cite{lo96} reanalized by us.
To compare our data with the RLFs in the literature, a bin $\Delta$\,log P 
$=$ 0.4 has been used.  Such a comparison is shown in Fig.~\ref{fig9}. 
We note that the comparison RLFs refer to R/R$_{A} \le$ 0.3 (the one by 
Fanti, which was computed within 1 R$_{A}$, has been rescaled using the 
radial distribution given in that paper) while the  NEP RLF is for 
R/R$_{A} \le$ 0.2. According to the radial distributions of cluster radio 
sources given in \cite{fan84} and \cite{lo95} a  multiplicative factor of 
$\sim 1.2 \pm 0.05$ would have to be applied to the NEP counts in order to 
make them comparable to the R/R$_{A} \le$ 0.3 radio counts,
thus even strengthening the conclusions of the  present study.

\begin{figure}
   \centering
  \includegraphics[bb=-10 100 574 574, width=12cm, height=11.5cm, 
                  angle=-90]{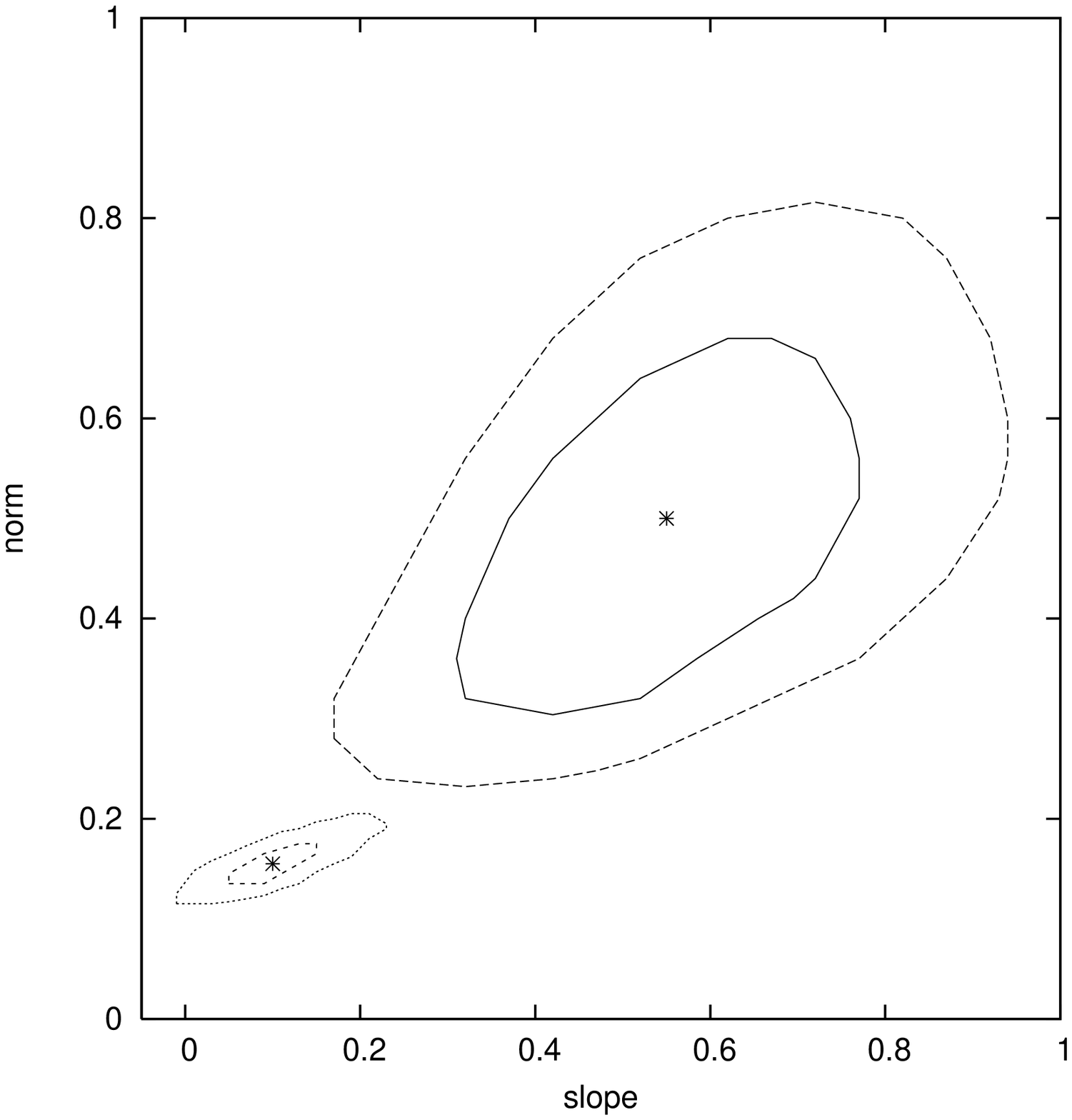}
   \vspace{+1cm}
   \caption{Confidence level contours  corresponding to 68\% and 95\% 
       probability for the parameters of Table~\ref{tab7} relative to the 
       radio powers below the break,  log P$_{1.4}$ (W Hz$^{-1}$) $\le$ 24.8.}
\label{fig10}
\end{figure}

\begin{figure}
   \centering
  \includegraphics[bb=0 100 574 574, width=13.7cm, height=10cm]{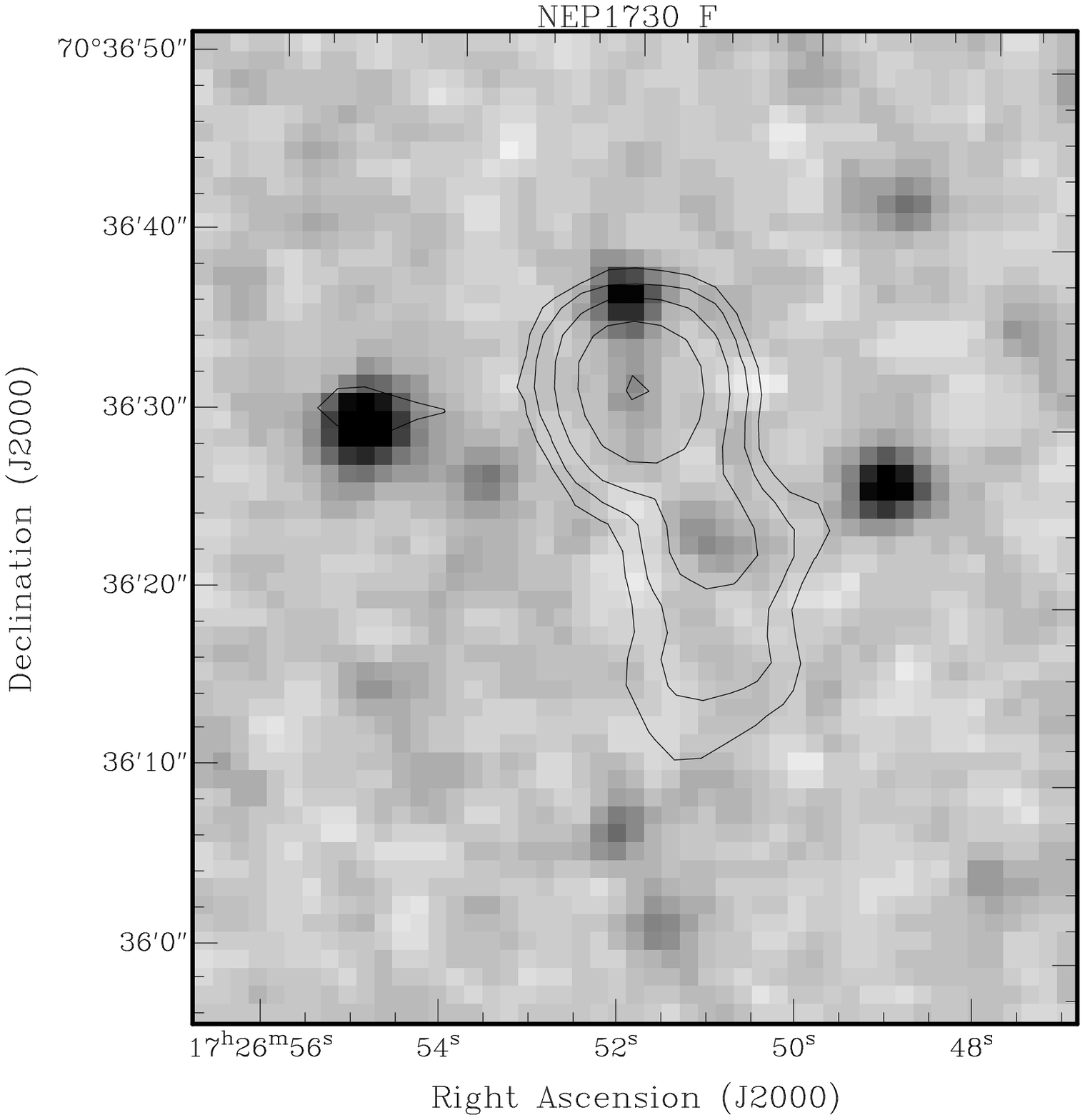}
   \includegraphics[bb=-5 100 574 544, width=14cm]{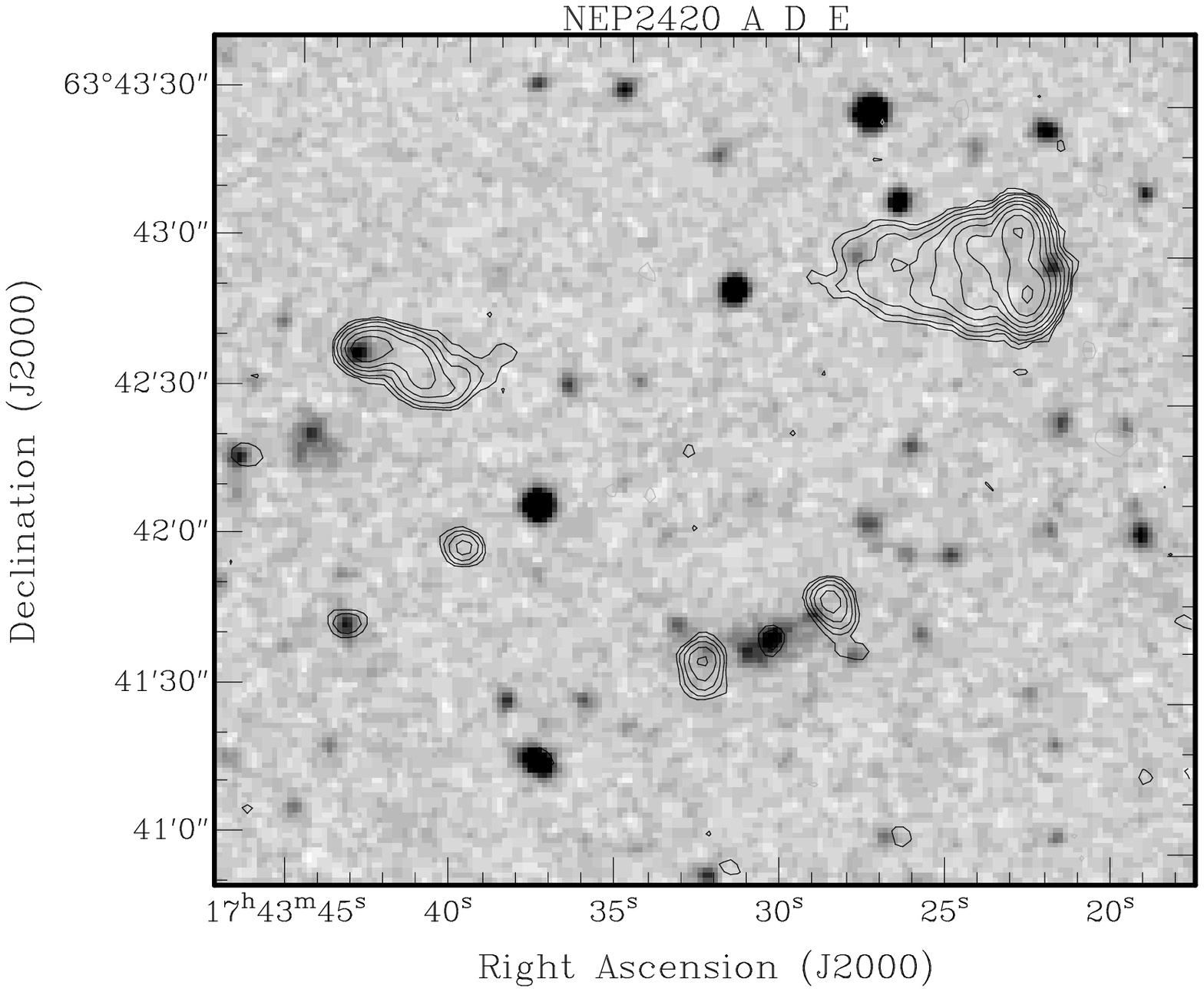}   
\vspace{1cm}
   \caption{VLA radio contour levels of the NEP sources ({\em top})
         NEP\,1730 F and ({\em bottom}) NEP\,2420 A (the straight triple 
         source associated to the cD to the South of the image), D 
         (the extended head-tail source to the North-West), E (the head-tail 
         source to the East), overplotted on the optical red Digitized Sky 
         Survey DSS\,II images. The first contour in each map is at the 
         3$\sigma$ noise level (see Table~\ref{tab2}) in \mJ and the following
         contours are at 2$^{n}$ this level.}
\label{fig11}
\end{figure}

\begin{figure}
\includegraphics[bb=-10 0 574 504, width=14cm]{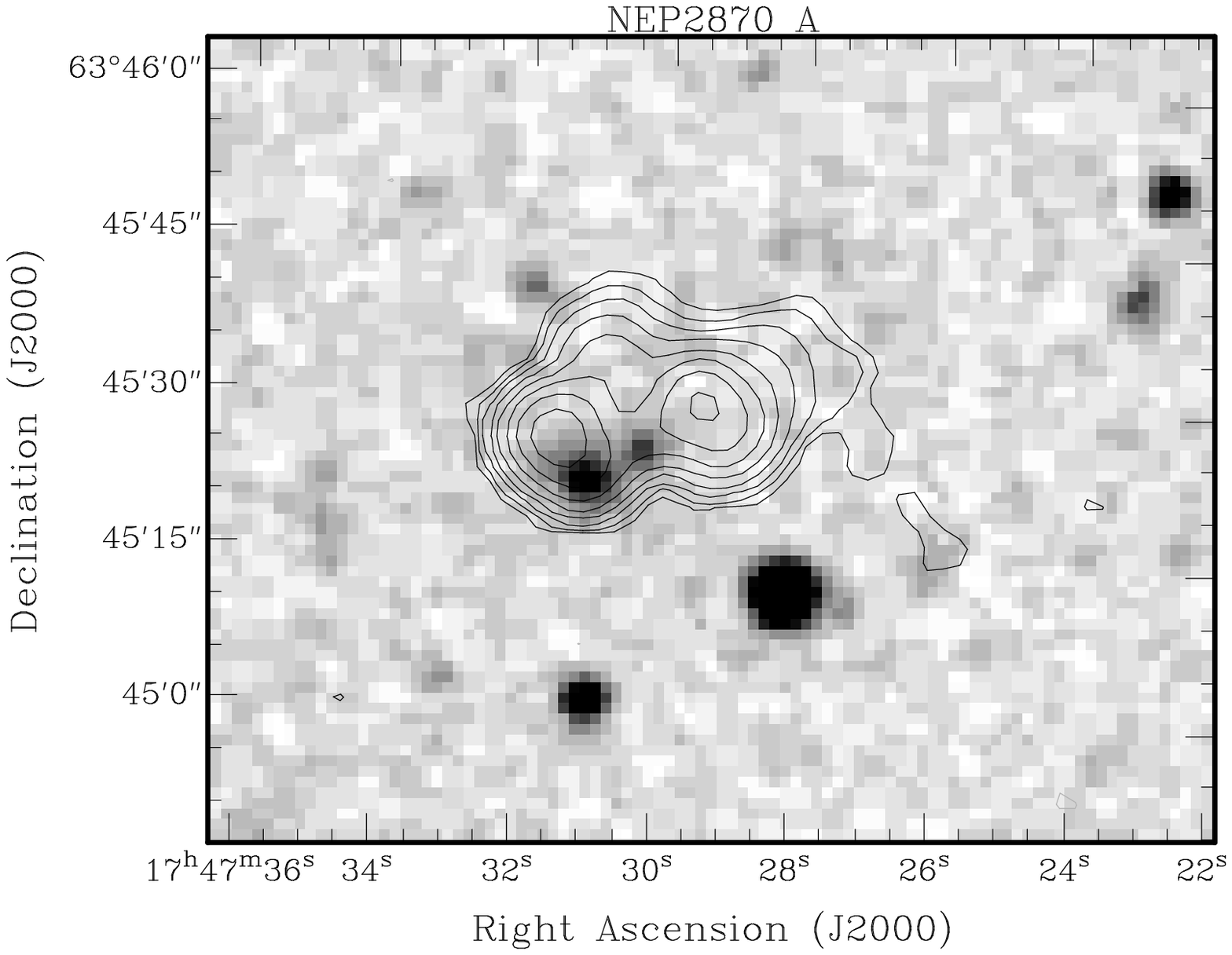}
\includegraphics[bb=10 30 574 454, width=14cm]{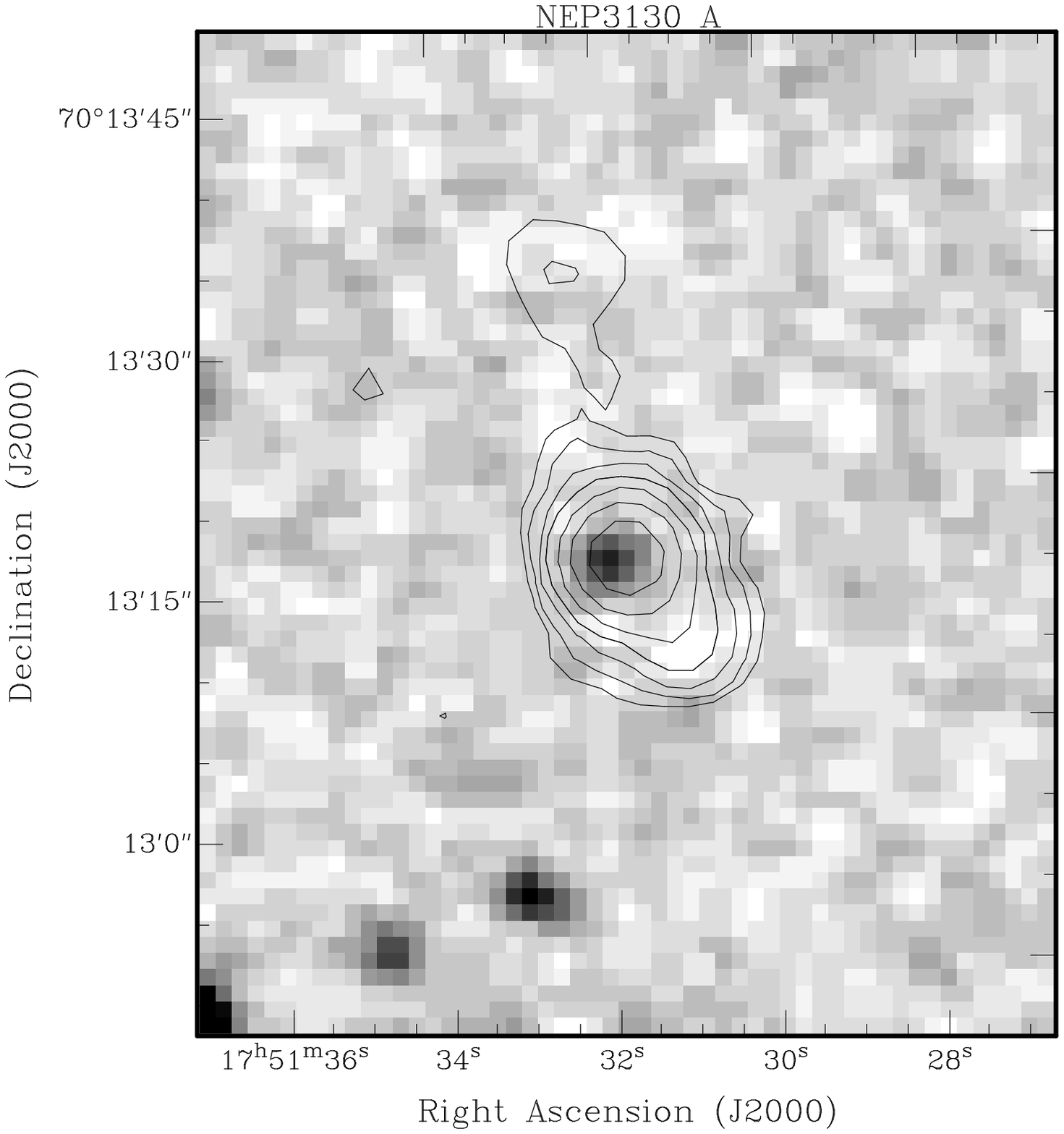}   
   \caption{VLA radio contour levels of the NEP sources ({\em top})
         NEP\,2870 A and ({\em bottom}) NEP\,3130 A  overplotted on 
         the optical red Digitized Sky Survey DSS\,II images. 
         The first contour in each map is at the 3$\sigma$ noise level
         (see Table~\ref{tab2}) in \mJ and the following contours
         are at 2$^{n}$ this level.}
\label{fig12}
\end{figure}

\begin{figure}
   \centering
  \includegraphics[bb=0 0 574 504, width=14cm]{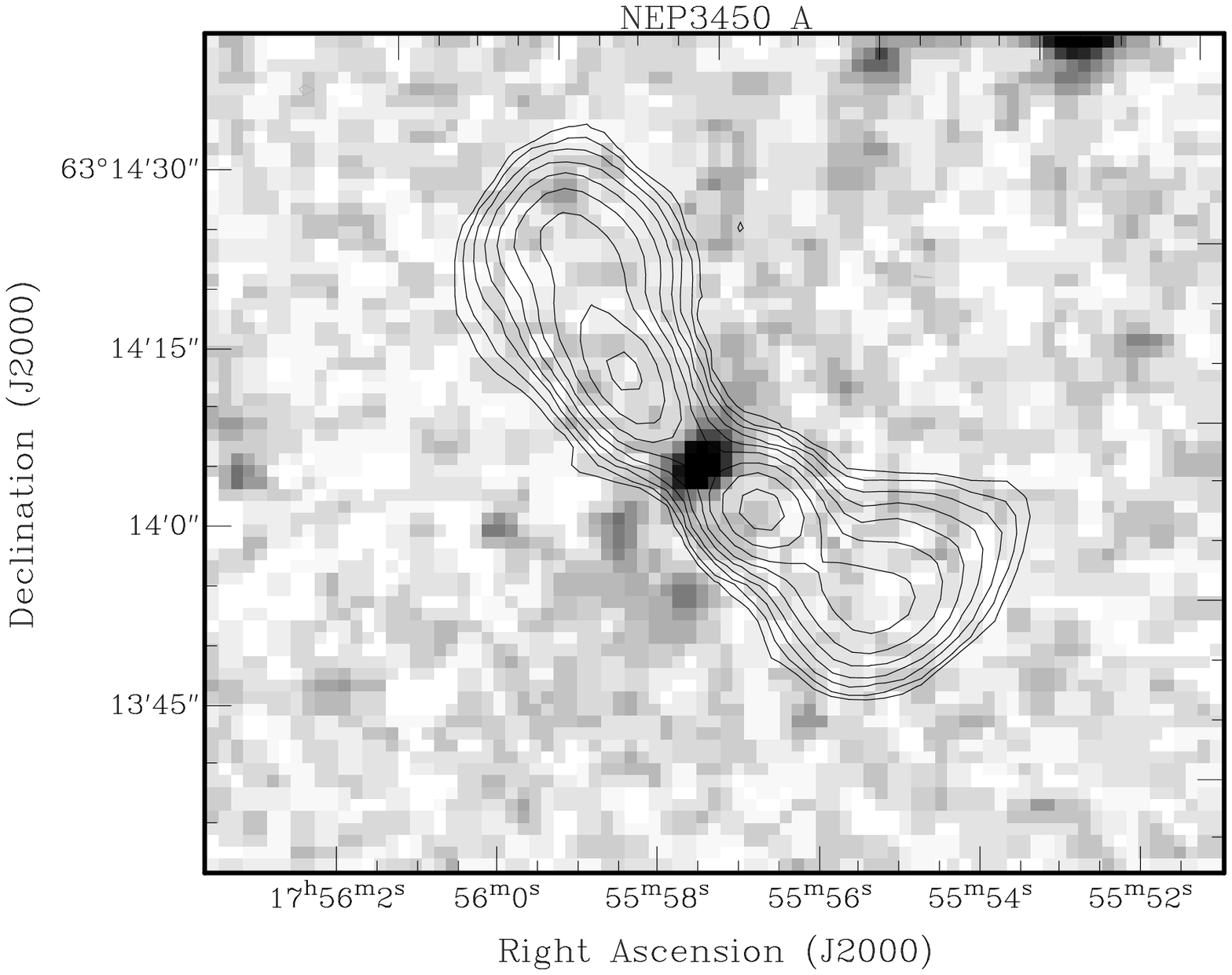}
  \includegraphics[bb=15 70 574 404, width=17cm]{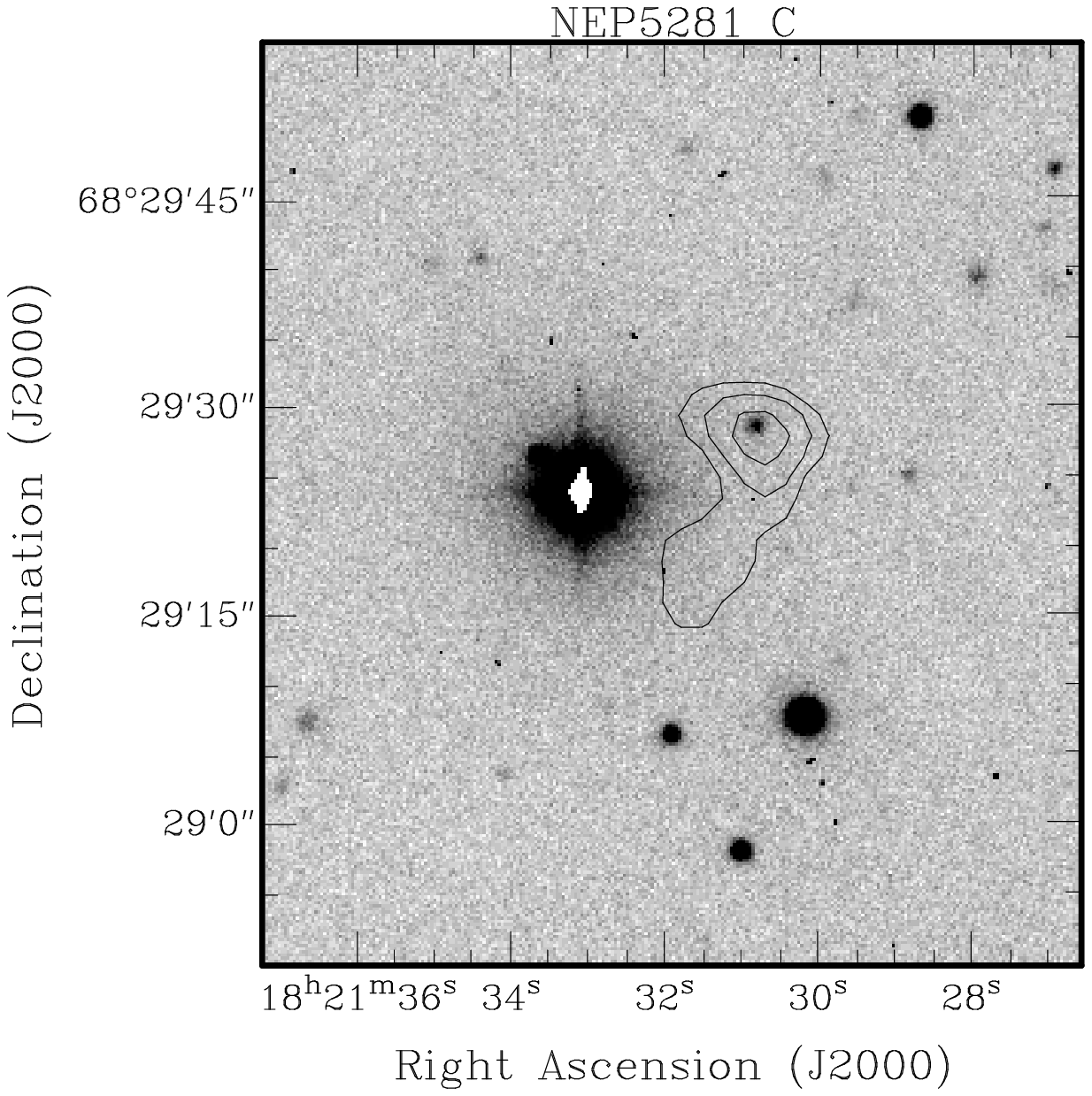}
   \caption{VLA radio contour levels of the NEP sources ({\em top})
       NEP\,3450 A overplotted on the optical red Digitized Sky 
         Survey DSS\,II image, and of ({\em bottom}) NEP\,5281 C 
         overplotted on a I-band UH 2.2m image.
         The first contour in each map is at the 3$\sigma$ noise level
         (see Table~\ref{tab2}) in \mJ and the following contours
         are at 2$^{n}$ this level.}
\label{fig13}
\end{figure}

\medskip
\noindent
As it can be seen from the figure, the NEP RLF lies above the local  RLFs.
The significance of this difference was checked by computing 
the  number of cluster radio galaxies expected in the NEP survey from a
RLF similar to the local ones. Down to log P$_{1.4}$ (W Hz$^{-1}$) $=$ 
22.5 we would have expected 8 cluster radio sources within  $R/R_A \le 0.2$ 
against  $\sim$22 (contamination corrected)  objects found. The difference 
is significant at 2.5 $\sigma$ level. In addition we have no evidence  of the 
break at $\approx$ log P$_{1.4}$ (W Hz$^{-1}$) $=$ 24.8 \cite[e.g.][]{lo96}. 
Actually, according to the \cite{lo96} RLF we would have  expected one 
source above the break, while we have four. In addition the slope 
of the NEP RLF appears steeper than the one found in the local samples.
To stress further these findings  we have applied the Maximum Likelihood 
Method  to the NEP and \cite{lo96} data.  Both datasets have been fitted 
with a power law of the form  $ F(P) = A \times P^{-\eta}$, where $F(P)$ is 
the number of sources per cluster in an interval $\Delta\log(P) = 0.4$ 
centered on $P$. The results of the fit are given in Table~\ref{tab7}.
The results  of the Ledlow and Owen data are the same as those given in 
\cite{lo96}. The Fig.~\ref{fig10} shows the areas of 1$\sigma$ and  
2$\sigma$ confidence level contours for the two fits.
%-------------------------------- Fig 10  ----------------------------------
%Fig. 10
%----------------------------------------------------------------------
The probability that the two RLFs are drawn from the same distribution is
$\le$0.25\% and even less if we take into account that the NEP RLF 
refers to a cluster area smaller than the one of the comparison sample. 
Finally, we note that the RLF by \cite{sto99} is higher than ours by a 
factor about two except in their lower radio power bin where the same 
authors suspect some incompleteness.

\medskip
\noindent
Although at present we cannot exclude evolutionary effects in the cluster
source population, we have to evaluate whether the differences we see between 
high and low redshift clusters  could be due to a selection effect. 
It is worth noting that the two high redshift cluster samples were X-ray 
selected and therefore they are overluminous in the X-ray band as compared to 
the low redshift samples. \cite{ab83} and earlier \cite{ba77} suggested  
that the X ray luminosity of a cluster is related to the central ($\le $
0.5 Mpc) galaxy density  approximately  as a square law ($ L_x \propto 
N_{gal}^{2}$). On this basis  \cite{sto99} computed a fractional 
RLF per galaxy assuming a higher galaxy population density \citep[see 
Figure 8 and  Table 8, column 3 in][]{sto99} and found no  differences with 
the one of Ledlow and Owen (1996).
The median X-ray luminosities of the \cite{lo96},  NEP  and \cite{sto99} 
samples are  $\sim 10^{43}$  \citep[see][]{lv03}, $\sim 6 \times 
10^{43}$  and $\sim 4.7 \times 10^{44}$ erg/sec respectively.
According to the scaling relations of \cite{ab83} and \cite{ba77}, we would
expect central galaxy population ratios of 1, 2.4 and 6.8. These ratios 
of galaxy populations would explain the differences in amplitude of the 
luminosity functions, even though \cite{riz03} find a weak anticorrelation 
in X-ray bright z$\sim$0.2  clusters between the  radio fraction and optical 
richness. A different or additional explanation is required to justify 
the steeper slope, and the absence of a break in the NEP RLF of cluster 
galaxies. We may speculate that different X-ray  cluster luminosities 
are related to different shapes of the optical luminosity
function of the cluster galaxies. A higher fraction of optically very bright 
galaxies could explain the apparent lack of the  break in the NEP RLF, as the 
break increases with optical luminosity \citep{lo96}. However, 
\cite{bes05} find no evidence for any dependence of the break luminosity 
on black hole mass by examining the properties of the host galaxies of a 
well defined sample of over 2000 radio  loud AGN from the Sloan Digital Sky 
Survey (SDSS). Thus this explanation could not be viable in our case. 
On the other hand at  lower radio luminosities a strongly  enhanced star 
formation activity  extending up to $10^{23}$ W Hz$^{-1}$ could explain the 
steeper slope at the fainter radio end. This evolution of the radio galaxy 
population as a function of redshift was already observed in z$\sim$0.4 
clusters  by \cite{mo99} who suggested that low-power radio sources are 
presumably  linked to an increase in the star formation. 
{\bb In our sample there is a fraction of low power radio galaxies with 
bluer SuperCOSMOS colors (3 out of 5 galaxies with log(P$_{I})<$23 have 
colors of order {\em B-R}$\sim$ 0.8--1.3) which might represent star-forming 
galaxies.  However the presence of broad emission lines in their optical 
spectra rules out the star formation scenario. In general the optical 
spectroscopic data available for most of the sample sources do not have 
the sufficient quality  to discriminate an AGN from a star forming galaxy.}
\vskip 0.2truecm
\noindent
If one adopts the currently-favored cosmological concordance model
\citep[$\Omega_{\Lambda}$ = 0.73, $\Omega_{m}$=0.27 and
H$_0=$ 71 km s$^{-1}$ Mpc$^{-1}$,][]{sper03} the derived  luminosities
of the NEP sources would have been 29\% higher at z $=$ 0.3 and 39\%
higher at z $=$ 0.8. Part of these percentages (12\%) is due to the
different Hubble costant adopted  and is redshift independent.
With the concordance cosmology the search area radius of 0.2 $R_A$ becomes
smaller by $\approx$ 10\% at z $=$ 0.8 and by $\approx$ 7\% at z $=$ 0.3.
The use of the concordance cosmology  does not affect the number
of sources used for the computation of the RLF but instead shifts the RLF  
towards higher radio luminosities, although not by a constant factor. The 
effect of the new cosmology  on the nearby cluster  sample is very small. 
The shift to higher luminosity due to the concordance model is mainly due 
to the different $H_0$ (12\%). So the difference found between distant and 
nearby cluster samples would be even slightly enhanced.

\vskip 0.2truecm
\noindent
Any interpretation of  these new  results has to await for more  data and 
information  both on the  radio side, in terms of better statistics, and on 
the optical side, in terms  of  complete spectroscopic and photometric data, 
galaxy classification, proper galaxy counts and  optical luminosity functions. 

\section{Summary and Conclusions}

We have used the complete NEP sample of distant galaxy clusters 
(0.3 $<$ z $<$ 0.8) as targets for a deep VLA continuum survey at 1.4 GHz 
down to 0.17 mJy beam$^{-1}$. The radio survey has discovered 79 radio 
sources within half Abell radius (1 Mpc in the cosmology adopted here) of 
which 32 are within 0.2  Abell radius. This radio sample is the most numerous 
sample of distant objects to compare with the detailed studies of radio 
galaxies in nearby Abell clusters conducted by Fanti (1984) and by Owen and 
collaborators \citep[among others][]{oo85,lo95,lo96,ol97,ol99,mo03}. The 
field source counts of the NEP radio sources are in agreement within the
errors with the VLA-VIRMOS deep field survey indicating that the NEP sample 
is  statistically complete.

\medskip
\noindent
Approximately  68\% of the NEP radio sources are unresolved, 16\% (13 sources)
have angular sizes in the range $\sim$8$''$ to $\sim$40$''$.
The extended  sources exhibit all kinds of morphology, from head--tail 
morphology, to  double sources and even one triple source. The identification 
of a radio source with a cluster  member has been made on the basis of the 
spectroscopic redshift (available only for 14 galaxies, belonging to 11 
different clusters), of the absolute magnitude of the associated object 
(see Section~\ref{optical}) and on its optical classification on the  
SuperCOSMOS/POSS\,II catalogue (see Section~\ref{opt-id}). Radio powers of the 
cluster sources range from log P$_{1.4}$ (W Hz$^{-1}$) $=$ 22.5 to 
log P$_{1.4}$ (W Hz$^{-1}$) $=$ 26.0. 

\medskip\noindent
The RLF for the NEP distant clusters has been computed as the number of radio 
galaxies per cluster as a function of radio power, and using only those 
radio galaxies within R/R$_{A}<$ 0.2 where field contamination is minimum. 
Previous radio observations  of X-ray distant galaxy clusters were performed 
by \cite{sto99} and  \cite{per03}. Neither study found any significant
evidence for evolution in the population  of distant cluster radio galaxies 
when compared to lower redshift clusters. The above mentioned authors
normalize  their RLF by a higher central galaxy population 
(according to the correlation between X-ray luminosity and galaxy 
counts of Abramopoulos \& Ku, 1983) when  comparing their RLF with the 
low-z cluster RLF. Thus it was surprising to discover  a change in the NEP 
RLF, both in slope and amplitude with respect to the local cluster samples, 
and the absence of a break at about log P$_{1.4}$  (W Hz$^{-1}$) $=$ 24.8 
observed  by \cite{fan84} and \cite{lo96}  in nearby rich Abell clusters. 

\medskip\noindent
Although it is tempting to claim a kind of density evolution in the RLF of 
distant clusters, coupled with a luminosity evolution at higher redshift, 
we believe that better statistics and sample complete optical identification
are required. Even if  evolutionary effects cannot be completely excluded, 
the higher amplitude of the NEP cluster RLF could be explained in terms of 
higher X-ray luminosity and thus higher central galaxy population density in 
the present sample with respect to the \cite{lo96} sample, as invoked by  
\cite{sto99}. We may consider other, {\em ad hoc}, effects to explain the 
absence of a break in the NEP RLF and the steepness of the slope. For 
instance a higher fraction of optically very bright galaxies would shift the 
break to higher radio powers \citep{lo96} (although Best et al., 2005, find 
no evidence for a dependence of the radio break with the black hole mass of 
the radio loud AGN hosted by the galaxies). On the other hand a strongly 
enhanced star formation activity at the faint radio end extending up to 
roughly log P (W Hz$^{-1}) =$ 23 would steepen the RLF slope  \citep{mo99}. 
In absence of more data these are only speculations awaiting to be confirmed 
by a much better statistics and deeper optical and radio data. 

\smallskip\noindent
The main result of this study is that the RLF of the distant X-ray clusters 
is very different from that of the local rich Abell clusters.

%-----------------------------   Fig 11  -----------------------------
%Fig 11
%--------------------------------------------------------------

%-----------------------------   Fig 12  -----------------------------
%Fig 12
%--------------------------------------------------------------
%-----------------------------   Fig 13  -----------------------------
%Fig 13
%--------------------------------------------------------------

\begin{acknowledgements}
This research uses data obtained with the National Radio Astronomy Observatory
Very Large Array: NRAO is operated by Associated Universities, Inc., under 
cooperative agreement with the National Science Foundation. The authors 
wish to thank an anonymous and competent referee for several helpful comments 
and suggestions.
\end{acknowledgements}

\clearpage
\bibliographystyle{aa}
\bibliography{bib}

\clearpage
%------------------------------Table 2-------------------------------------
\tabcolsep 0.28cm
\begin{landscape}
\begin{longtable}{lcccccrrcccrr}
\caption{The Radio Source Catalogue}
\\
\hline
\hline
Cluster &  rms  &Source&  RA(J2000)& DEC(J2000)& R/R$_A$ &  S$^{corr}_P$ & 
S$_I$ & $\Theta_M \times \Theta_m$ &  P.A.& Notes &\\
       & \mJ & & & & & \mJ & mJy & \arcsec  & \deg &  &\\
\hline
\endhead
NEP\,200 & 0.034 & A & 17 57 21.36 & 66 29 45.1&  0.32 &   0.43$\pm$0.03 
& 0.43$\pm$0.03 &	        &	& u \\
\hline
NEP\,310 & 0.033 & A & 17 59 03.58 & 65 20 02.5&  0.17 &   0.94$\pm$0.04 &
1.06$\pm$0.06 &  2.6 & 166.2& ru\\  
       &       & B   & 17 59 05.07 & 65 20 57.8&  0.12 &   0.30$\pm$0.03 & 
       0.30$\pm$0.03 &	  	 &	& u \\  
       &       & C   & 17 59 05.67 & 65 18 47.9&  0.33 &   0.34$\pm$0.03 & 
       0.34$\pm$0.03 &	   	 &	& u  \\ 
       &       & D   & 17 59 28.67 & 65 21 41.5&  0.45 &   0.41$\pm$0.03 & 
       0.41$\pm$0.03 &	   	 &	& u  \\ 
\hline 
NEP\,1730& 0.032 & A & 17 27 36.49 & 70 35 23.4&  0.05 &   0.42$\pm$0.03 & 
0.58$\pm$0.06 &  4.4  & 97.9 & ru \\ 
       &       & B   & 17 27 33.60 & 70 35 47.0&  0.00 &   0.26$\pm$0.03 & 
       0.78$\pm$0.14 & 8 	 &35	& ext \\ 
       &       & C   & 17 27 57.97 & 70 34 54.6&  0.26 &   0.31$\pm$0.03 & 
       0.31$\pm$0.03 &	   	 &	& u\\  
       &       & D   & 17 27 57.38 & 70 35 40.9&  0.23 &   0.19$\pm$0.04 & 
       0.36$\pm$0.06 &	5   	 &7	& ext\\ 
       &       & E   & 17 28 16.73 & 70 36 34.9&  0.43 &   0.24$\pm$0.03 & 
       0.24$\pm$0.03 &	   	 &	& u\\  
       &       & F   & 17 26 52.03 & 70 36 31.7&  0.42 &   2.20$\pm$0.08 & 
       5.21$\pm$0.21 &	18   	 &17	& ext\\ 
\hline
NEP\,1780& 0.033 & A & 17 28 38.38 & 70 41 02.6&  0.00 &   4.84$\pm$0.15 & 
       4.84$\pm$0.15 &           &      & u\\  
       &       & B   & 17 28 55.74 & 70 40 11.4&  0.28 &   4.30$\pm$0.13 & 
       4.30$\pm$0.13 &            &      & u\\ 
       &       & C   & 17 28 34.06 & 70 39 09.1&  0.32 &   0.20$\pm$0.02 & 
       0.20$\pm$0.02 &	   	 &	& u \\ 
       &       & D   & 17 28 09.03 & 70 41 04.1&  0.39 &   0.27$\pm$0.03 & 
       0.27$\pm$0.03 &	   	 &	& u \\ 
\hline
NEP\,2420& 0.032 & A & 17 43 30.47 & 63 41 41.3&  0.00 &   2.53$\pm$0.08 &
6.42$\pm$0.18 &	28.4   	 &117 & ext \\
       &  & B        & 17 43 39.84 & 63 41 58.4&  0.13 &   1.11$\pm$0.04 & 
       1.11$\pm$0.04 &	   	 &	& u\\ 
       &       & C   & 17 43 43.31 & 63 41 42.6&  0.18 &   0.41$\pm$0.03 & 
       0.41$\pm$0.03 &	   	 &	& u\\ 
       &       & D   & 17 43 23.14 & 63 42 58.5&  0.19 &  29.55$\pm$0.89 &
       187.05$\pm$6.07&  36 x 14        & 96	& ext\\ 
       &       & E   & 17 43 43.04 & 63 42 38.0&  0.21 &   6.75$\pm$0.21 & 
       21.14$\pm$0.66&  21         &69	& ext\\ 
       &       & F   & 17 44 03.11 & 63 41 16.8&  0.45 &   0.84$\pm$0.05 & 
       4.43$\pm$0.20&  16  &13	& ext\\ 
       &       & G   & 17 44 05.82 & 63 41 16.4&  0.49 &   0.43$\pm$0.03 & 
       0.43$\pm$0.03 &	   	 &	& u\\ 
       &       & H   & 17 44 03.99 & 63 40 12.3&  0.49 &   0.32$\pm$0.05 & 
       1.63$\pm$0.13 & 9     	 & 120	& ext\\ 
       &       & I   & 17 43 37.39 & 63 41 16.3&  0.11 &   0.21$\pm$0.02 & 
       0.21$\pm$0.02 &	   	 &	& u\\ 
       &       & L   & 17 43 46.43 & 63 42 16.1&  0.23 &   0.23$\pm$0.03 & 
       0.23$\pm$0.03 &	   	 &	& u\\ 
       &       & M   & 17 43 39.55 & 63 43 49.0&  0.29 &   0.20$\pm$0.03 & 
       0.20$\pm$0.03 &	   	 &	& u\\ 
\hline
NEP\,2560& 0.028 & A & 17 45 17.07 & 65 53 34.3&  0.37 &   0.25$\pm$0.02 & 
       0.25$\pm$0.02 &	     	 &	& u \\ 
       &       & B   & 17 45 16.19 & 65 56 37.3&  0.16 &   0.25$\pm$0.02 & 
       0.25$\pm$0.02 &	   	 &	& u \\ 
       &       & C   & 17 45 32.84 & 65 57 54.8&  0.46 &   0.31$\pm$0.02 & 
       0.31$\pm$0.02 &	   	 &	& u \\ 
       &       & D   & 17 45 27.02 & 65 58 02.4&  0.43 &   0.30$\pm$0.03 & 
       0.30$\pm$0.03 &	   	 &	& u  \\ 
\hline				       
NEP\,2770& 0.031 & A & 17 46 46.30 & 66 39 02.5&  0.01 &   0.49$\pm$0.03 & 
       1.05$\pm$0.09 &	10   	 &10	& ext\\ 
       &       & B   & 17 46 53.41 & 66 40 21.9&  0.20 &   0.69$\pm$0.04 & 
       0.87$\pm$0.06 &  3.7 & 172.4& ru\\  
       &       & C   & 17 46 54.88 & 66 39 19.7&  0.11 &   0.20$\pm$0.02 & 
       0.20$\pm$0.02 &            &	 & u\\   
\hline
NEP\,2870& 0.027 & A & 17 47 30.58 & 63 45 27.2&  0.01 &  17.86$\pm$0.54 &
       74.44$\pm$2.24 & 29          &115	& ext\\  
       &       & B   & 17 47 42.40 & 63 45 05.6&  0.16 &   3.48$\pm$0.11 & 
       6.18$\pm$0.24 &  6.7  & 38.8 & ru\\   
       &       & C   & 17 47 12.83 & 63 47 08.9&  0.33 &   0.63$\pm$0.03 & 
       0.63$\pm$0.03 &            &      & u\\ 
       &       & D   & 17 47 01.15 & 63 44 46.0&  0.42 &   9.63$\pm$0.29 & 
       9.63$\pm$0.29 &            &      & u\\ 
       &       & E   & 17 47 41.76 & 63 41 57.8&  0.45 &   0.46$\pm$0.03 & 
       0.46$\pm$0.03 &            &      & u\\ 
       &       & F   & 17 47 34.28 & 63 43 20.7&  0.25 &   0.26$\pm$0.02 & 
       0.26$\pm$0.02 &	   	 &	& u \\ 
       &       & G   & 17 47 44.35 & 63 48 18.2&  0.40 &   0.26$\pm$0.03 & 
       0.26$\pm$0.03 &	   	 &	& u \\ 
       &       & H   & 17 47 46.49 & 63 48 24.6&  0.43 &   0.29$\pm$0.03 & 
       0.29$\pm$0.03 &	         &	& u \\ 
\hline
NEP\,2950& 0.030 & A & 17 48 57.56 & 70 20 02.1&  0.21 &   1.23$\pm$0.07 & 
       1.23$\pm$0.04 &	   	 &	& u  \\ 
       &       & B   & 17 48 23.03 & 70 20 17.4&  0.18 &   0.36$\pm$0.03 & 
       0.36$\pm$0.02 &	   	 &	& u\\   
       &       & C   & 17 48 13.51 & 70 19 16.8&  0.33 &   0.44$\pm$0.04 & 
       0.79$\pm$0.10 &	9   	 &45	& ext\\ 
       &       & D   & 17 48 06.07 & 70 21 25.3&  0.37 &   0.69$\pm$0.04 & 
       0.68$\pm$0.03 &	   	 &	& u \\ 
       &       & E   & 17 48 41.91 & 70 24 13.2&  0.45 &   0.99$\pm$0.06 & 
       0.99$\pm$0.04 &            &      & u\\ 
\hline
NEP\,3130& 0.052 & A & 17 51 32.69 & 70 13 21.3&  0.00 &   18.35$\pm$0.55&
       24.39$\pm$0.77 &       8.1   & 37   & ext\\
       &       & B   & 17 51 22.24 & 70 13 19.3&  0.14 &   1.21$\pm$0.05 &
       1.21$\pm$0.05&           &      &  u  \\
       &       & C   & 17 51 31.09 &  70 15 01.2&  0.26 &   3.48$\pm$0.12 &
       3.76$\pm$0.15& 2.29         &      & ru \\      
\hline 
NEP\,3200& 0.028 & A & 17 52 12.94 & 65 22 36.7&  0.08 &   0.19$\pm$0.02 & 
       0.19$\pm$0.02 &	   	 &	& u  \\ 
       &       & B   & 17 52 01.03 & 65 22 56.9&  0.10 &   0.94$\pm$0.04 & 
       1.24$\pm$0.06 &  3.9  & 148.0& ru\\ 
       &       & C   & 17 52 30.71 & 65 22 32.2&  0.33 &   2.19$\pm$0.07 & 
       2.19$\pm$0.07 &            &      & u\\ 
\hline
NEP\,3320& 0.031 & A & 17 54 37.87 & 69 06 05.2&  0.16 &   0.23$\pm$0.02 & 
       0.23$\pm$0.02 &	   	 &	& u \\ 
       &       & B   & 17 54 09.36 & 69 03 01.0&  0.48 &   6.18$\pm$0.19 & 
       6.18$\pm$0.19 &            &     & u\\  
\hline
NEP\,3450& 0.032 & A & 17 55 57.90 & 63 14 08.9&  0.00 &  35.22$\pm$1.06 &
       183.44$\pm$5.51 & 40         &48	& ext\\ 
       &       & B   & 17 55 44.76 & 63 12 38.9&  0.29 &   0.60$\pm$0.04 & 
       0.74$\pm$0.07 &  3.6  & 77.6 & ru\\ 
\hline
NEP\,4150& 0.031 & A & 18 05 57.23 & 68 13 27.7&  0.08 &   0.32$\pm$0.03 & 
       0.32$\pm$0.03 &	   	 &	& u \\ 
       &       & B   & 18 06 04.54 & 68 10 48.8&  0.29 &   0.71$\pm$0.03 & 
       0.71$\pm$0.03 &	   	 &	& u\\  
       &       & C   & 18 06 22.66 & 68 11 22.6&  0.29 &   0.33$\pm$0.03 & 
       0.33$\pm$0.03 &	   	 &	& u \\ 
       &       & D   & 18 05 42.68 & 68 13 22.6&  0.24 &   0.73$\pm$0.03 & 
       0.73$\pm$0.03 &	   	 &	& u \\ 
       &       & E   & 18 05 30.88 & 68 12 52.3&  0.37 &   4.90$\pm$0.15 & 
       6.67$\pm$0.22 &  5.6  & 106.8& ru \\ 
       &       & F   & 18 05 30.05 & 68 10 40.5&  0.49 &   0.34$\pm$0.04 & 
       0.34$\pm$0.04 &	   	 &	& u\\ 
       &       & G   & 18 06 12.74 & 68 15 58.8&  0.33 &   0.26$\pm$0.03 & 
       0.26$\pm$0.03&	   	 &	& u\\ 
       &       & H   & 18 06 04.78 & 68 13 16.3&  0.00 &   0.20$\pm$0.02 & 
       0.20$\pm$0.02 &	   	 &	& u \\ 
       &       & I   & 18 06 08.81 & 68 13 10.0&  0.04 &   0.21$\pm$0.02 & 
       0.21$\pm$0.02 &	   	 &	& u\\ 
       &       & L   & 18 05 54.48 & 68 14 19.5&  0.17 &   0.18$\pm$0.02 & 
       0.18$\pm$0.02 &	   	 &	& u\\ 
       &       & M   & 18 06 31.77 & 68 13 28.0&  0.29 &   0.22$\pm$0.03 & 
       0.22$\pm$0.03 &	   	 &	& u\\ 
\hline
NEP\,4560& 0.029 & A  & 18 11 16.55 & 64 47 22.4&  0.04 &   0.77$\pm$0.04 & 
       0.95$\pm$0.06 &  3.3  &140.42& ru \\ 
       &       & B   & 18 11 42.74 & 64 48 56.4&  0.44 &   0.44$\pm$0.03 & 
       0.44$\pm$0.03 &	   	 &	& u\\  
       &       & C   & 18 11 21.40 & 64 50 31.3&  0.47 &   0.24$\pm$0.03 & 
       0.24$\pm$0.03 &	   	 &	& u \\ 
       &       & D   & 18 11 09.36 & 64 49 00.6&  0.29 &   0.21$\pm$0.02 & 
       0.21$\pm$0.02 &	   	 &	& u \\ 
\hline
NEP\,4610& 0.026 & A & 18 12 08.18 & 63 53 31.8&  0.00 &   1.67$\pm$0.06 & 
       2.15$\pm$0.10 &  3.2 x 1.3 & 180.0& r \\ 
       &       & B   & 18 12 19.22 & 63 52 33.9&  0.25 &   0.32$\pm$0.03 & 
       0.32$\pm$0.03 &            &      & u\\ 
       &       & C   & 18 12 12.86 & 63 53 58.5&  0.11 &   0.22$\pm$0.02 & 
       0.22$\pm$0.02 &	   	 &	& u \\ 
       &       & D   & 18 11 55.59 & 63 51 25.2&  0.41 &   0.21$\pm$0.02 & 
       0.21$\pm$0.02 &	   	 &	& u \\ 
\hline
NEP\,5281& 0.032 & A & 18 21 37.12 & 68 27 50.9&  0.07 &   0.73$\pm$0.04 & 
       1.03$\pm$0.08 &  3.4 x 2.44& 32.3 & r\\ 
       &       & B   & 18 21 33.10 & 68 27 55.4&  0.00 &   0.61$\pm$0.03 & 
       0.61$\pm$0.03 &	   	 &	& u\\  
       &       & C   & 18 21 30.72 & 68 29 28.0&  0.30 &   0.60$\pm$0.04 & 
       1.14$\pm$0.10 &	12   	 &118	& ext\\ 
       &       & D   & 18 21 15.90 & 68 29 18.0&  0.40 &   0.36$\pm$0.03 & 
       0.36 $\pm$0.03 &	   	 &	& u\\       

\hline
\label{tab2}
\end{longtable}
\end{landscape}

\clearpage
%------------------------------Table 3-----------------------------------------
\tabcolsep 0.25cm
\begin{longtable}{lcccrrcccc}
\caption{Optical Counterparts of the Detected Radio Sources}
\\
\hline
\hline 
Cluster & Source&RA(J2000) & DEC(J2000) & $\Delta$RA & $\Delta$DEC & m$_R$ & m$_B$& Opt. & z  \\
 &   &  optical  &  optical  & (o-r) &  (o-r)  &        &      & Class &     \\ 
\hline
\endhead
NEP\,310 & B & 17 59 05.19 & 65 20 57.4&  + 0.7 & -- 0.4 & 18.7 & 20.7 & G & \\
\hline
\\
NEP\,1730& A & 17 27 36.36 & 70 35 23.6& -- 0.6 &  + 0.2 & 18.7 & 21.1 & G & \\
         & B & 17 27 33.47 & 70 35 47.7& -- 0.6 &  + 0.7 & 18.4 & 21.2 & G & 0.306* \\
         & C & 17 27 58.01 & 70 34 54.3&  + 0.2 & -- 0.3 & 19.7 & 20.7 & S & \\
         & E & 17 28 17.04 & 70 36 35.0&  + 1.5 &  + 0.1 & 20.1 & 22.5 & G & \\
         & F & 17 26 51.72 & 70 36 31.8& -- 1.5 &  + 0.1 & 20.6 &  --  & G & \\
\hline
\\
NEP\,1780& A & 17 28 38.40 & 70 41 02.7&  + 0.1 &  + 0.1 & 19.9 & 20.9 & S & 0.550* \\
         & C & 17 28 33.96 & 70 39 08.9& -- 0.5 & -- 0.2 & 20.0 & 21.5 & G & \\
         & D & 17 28 09.17 & 70 41 03.5&  + 0.7 & -- 0.6 & 20.7 & 21.8 & S & \\
\hline
\\
NEP\,2420& A & 17 43 30.47 & 63 41 41.4&    0.0 &  + 0.1 & 17.6 & 20.0 & G & 0.331* \\ 
         & C & 17 43 43.32 & 63 41 42.3&  + 0.1 & -- 0.3 & 18.8 & 20.5 & G & \\ 
         & D & 17 43 22.16 & 63 42 57.8& -- 6.5 & -- 0.7 & 18.9 & 21.5 & G & \\
         & E & 17 43 43.14 & 63 42 37.4&  + 0.7 & -- 0.6 & 18.5 & 21.2 & G & \\
         & F & 17 44 04.01 & 63 41 18.3&  + 6.0 &  + 1.5 & 18.5 & 21.1 & G & \\
         & G & 17 44 05.94 & 63 41 14.9&  + 0.8 & -- 1.5 & 17.2 & 19.2 & S & \\
         & I & 17 43 37.43 & 63 41 15.9&  + 0.3 & -- 0.4 & 17.3 & 18.1 & G & \\
         & L & 17 43 46.74 & 63 42 15.9&  + 2.1 & -- 0.2 & 18.8 & 20.3 & G & \\
\hline
\\
NEP\,2560& A & 17 45 17.06 & 65 53 34.9&    0.1 &  + 0.6 & 19.8 & 22.3 & G & \\
         & D & 17 45 26.98 & 65 58 02.5& -- 0.2 &  + 0.1 & 19.3 & 21.0 & G & \\
\hline
\\
NEP\,2770& A & 17 46 46.74 & 66 39 04.9&  + 2.6 &  + 2.4 & 18.5 & 22.0 & G & 0.392* \\
         & B & 17 46 53.63 & 66 40 22.2&  + 1.3 &  + 0.3 & 20.5 & 22.3 & S &	   \\ 
         & C & 17 46 54.90 & 66 39 20.4&  + 0.1 &  + 0.7 & 18.6 & 21.7 & G & 0.384* \\ 
\hline
\\
NEP\,2870& A & 17 47 30.26 & 63 45 25.5& -- 2.1 & -- 1.7 & 19.7 & 21.1
&S&          \\ 
         & B & 17 47 42.06 & 63 45 04.2& -- 2.2 & -- 1.4 & 18.1 & 20.8
& G&0.326*   \\ 
         & E & 17 47 41.80 & 63 41 57.6&  + 0.3 & -- 0.2 & 18.5 & 21.2
&G&	    \\ 
         & F & 17 47 34.26 & 63 43 20.0& -- 0.1 & -- 0.7 & 18.9 & 20.3
&G&	    \\
         & H & 17 47 46.45 & 63 48 25.4& -- 0.3 &  + 0.8 & 20.2 &  --  & G & \\ 
\hline
\\
NEP\,2950& C & 17 48 12.83 & 70 19 13.3& -- 3.4 & -- 3.5 & 20.5 &  --
&S&  	    \\
         & E & 17 48 41.95 & 70 24 12.9&  + 0.2 & -- 0.3 & 20.9 &  --  & G & \\
\hline
\\
NEP\,3130& A & 17 51 32.57 & 70 13 21.6& -- 0.6 &  + 0.3 & 19.8 &  --  & G & 0.491* \\
\hline
\\
NEP\,3200& A & 17 52 12.97 & 65 22 36.0&  + 0.2 & -- 0.7 & 18.6 & 19.4 & G & 0.394* \\ 
\hline
\\
NEP\,3450& A & 17 55 57.84 & 63 14 08.9& -- 0.4 &    0.0 & 18.5 & 21.7 & G & 0.386* \\
         & B & 17 55 44.73 & 63 12 38.6& -- 0.2 & -- 0.3 & 17.5 & 18.8 & G & \\ 
\hline
\\
NEP\,4150& F & 18 05 30.09 & 68 10 40.0&  + 0.2 & -- 0.5 & 18.4 & 20.6 & G & \\
         & H & 18 06 04.91 & 68 13 16.2&  + 0.7 & -- 0.1 & 19.0 & 20.3 & G & 0.295* \\ 
         & I & 18 06 08.99 & 68 13 09.6&  + 1.0 & -- 0.4 & 18.5 & 19.3 & S & \\
\hline
\\
NEP\,4560& A & 18 11 16.57 & 64 47 22.3&  + 0.1 & -- 0.1 & 16.9 & 18.9 & G & 0.174 \\ 
         & D & 18 11 09.31 & 64 49 01.2& -- 0.3 &  + 0.6 & 20.1 &  --  & S & \\
\hline
\\
NEP\,4610& A & 18 12 08.15 & 63 53 31.5& -- 0.2 & -- 0.3 & 20.4 & 21.8 & S & 0.545* \\ 
         & B & 18 12 19.39 & 63 52 34.3&  + 1.1 &  + 0.4 & 17.9 & 19.5 & G & \\ 
         & C & 18 12 12.82 & 63 53 59.0& -- 0.3 &  + 0.5 & 19.6 & 20.9 & G & \\ 
\hline
\\
NEP\,5281& A & 18 21 37.20 & 68 27 51.0&  + 0.4 &  + 0.1 &  --  &  -- &  & 0.815*  \\ 
         & B & 18 21 32.90 & 68 27 55.0& -- 1.1 & -- 0.4 &  --  &  -- &  & 0.817*  \\ 
         & C & 18 21 30.80 & 68 29 28.8&  + 0.4 &  + 0.8 &  --  &  -- &  & 0.820* \\
\hline
\label{tab3}
\end{longtable}
%------------------------------------------------------------------------------

\clearpage
%--------------------------------Table 4----------------------------------
\tabcolsep 0.2cm
\begin{table}[htb]
\begin{center}
\caption{The Sample of Radio Sources Used to Derive the RLF}
\begin{tabular}{lrccrrccc}
\hline 
\hline
Cluster &  Source  & log P$_{peak}$ (W Hz$^{-1}$ beam$^{-1}$)  & log P$_{I}$ (W Hz$^{-1}$)& $M_r$ & LLS (kpc) & W$_{RLF}$ & Comments$^{\mathrm{a}}$ \\
\hline
 NEP\,310  & A	 &  23.48   &	 23.53  & $>$-20.8 &  11.4 & 0.062  &  3 \\
          & B	 &  22.98   &	 22.98  &  -22.9   &       & 0.100  &  2 \\   
 NEP\,1730 & A	 &  22.97   &	 23.11  &  -22.3   &  17.3 & 0.100  &  2 \\   
          & B	 &  22.76   &	 23.23  &  -22.6   &  31.5 & 0.167  &  1 \\  
 NEP\,1780 & A	 &  24.59   &	 24.59  &  -23.3   &       & 0.059  &  1 \\
 NEP\,2420 & A	 &  23.81   &	 24.21  &  -23.6   & 116.8 & 0.059  &  1 \\
          & B	 &  23.45   &	 23.45  & $>$-20.5 &       & 0.062  &  3 \\
          & C	 &  23.01   &	 23.01  &  -22.5   &       & 0.100  &  2 \\   
          & D	 &  24.87   &	 25.67  &  -22.3   &  148  & 0.059  &  2 \\
          & I	 &  22.72   &	 22.72  &  -23.9   &       & 0.200  &  2 \\   
 NEP\,2560 & B	 &  23.41   &	 23.41  & $>$-22.9 &       & 0.062  &  2 \\   
 NEP\,2770 & A	 &  23.25   &	 23.58  &  -23.2   & 45.5  & 0.067  &  1 \\
          & B	 &  23.40   &	 23.50  &  -21.3   & 16.8  & 0.062  &  4 \\   
          & C	 &  22.87   &	 22.87  &  -23.1   &       & 0.111  &  1 \\
 NEP\,2870 & A	 &  24.66   &	 25.28  &  -21.6   & 119.5 & 0.059  &  4 \\   
          & B	 &  23.95   &	 24.20  &  -23.2   & 27.6  & 0.059  &  1 \\
 NEP\,2950 & B	 &  23.01   &	 23.01  & $>$-20.6 &       & 0.100  &  3 \\
 NEP\,3130 & A   &  25.06   &    25.19  &  - 22.9  &  42.1 & 0.059  &  1 \\
           & B   &  23.88   &    23.88  & $>$-22.7 &       & 0.059  &  2 \\
 NEP\,3200 & A	 &  22.85   &	 22.85  &  -23.2   &       & 0.111  &  1 \\
           & B	 &  23.55   &	 23.67  & $>$-21.0 &  18.0 & 0.059  &  3 \\
 NEP\,3320 & A	 &  23.19   &	 23.19  & $>$-22.0 &       & 0.077  &  2 \\
 NEP\,3450 & A	 &  25.11   &	 25.82  &  -23.2   & 181.6 & 0.059  &  1 \\
 NEP\,4150 & A	 &  22.84   &	 22.84  & $>$-20.2 &       & 0.111  &  3 \\
          & H	 &  22.63   &	 22.63  &  -21.9   &       & 0.333  &  1 \\
          & I	 &  22.64   &    22.64  &  -22.5   &       & 0.250  &  4 \\ 
          & L	 &  22.58   &	 22.58  & $>$-20.2 &       & 0.500  &  3 \\
 NEP\.4560 & A   &  23.59   &    23.69  &  -22.6$^{\mathrm{b}}$ &  8.7$^{\mathrm{b}}$ & 0.059 &  5 \\ 
 NEP\,4610 & A	 &  24.12   &	 24.22  &  -22.7   & 17.4  & 0.059  &  1 \\
          & C	 &  23.24   &	 23.24  &  -23.5   &       & 0.071  &  2 \\    
 NEP\,5281 & A	 &  24.17   &	 24.32  & $>$-24.4 & 21.8  & 0.059  &  1 \\
          & B	 &  24.09   &	 24.09  & $>$-24.4 &       & 0.059  &  1 \\
\hline
\label{tab4}
\end{tabular}
\begin{list}{}{}
\item[$^{\mathrm{a}}$] 1 $=$ cluster member, radio galaxy has a measured 
redshift;  2 $=$ assumed to be a cluster member, radio galaxy satisfies 
the optical magnitude criterion described in Section~\ref{optical};
3 $=$ possible field source, radio galaxy is a possible contaminant according 
to the optical magnitude criterion described in Section~\ref{optical};
4 $=$ star-like object, possible non cluster member; 5 $=$ foreground galaxy:
NEP\,4560 A.
\item[$^{\mathrm{b}}$] The absolute red magnitude and the large 
linear size are calculated using the spectrcopic redshift 0.174.  
\end{list}
\end{center}
\end{table}

\clearpage
%------------------------------Table 5 --------------------------------------
\tabcolsep 0.8cm
\begin{table}[htb]
\begin{center}
\caption{The NEP Differential and Integral RLFs (Contaminant Subtraction 
Method)}
\begin{tabular}{c|cc}
\\
\hline\hline
          & Differential  & Integral \\
\hline
Log P$_{1.4}$ $^{\mathrm{a}}$ & N/cluster        & N/cluster \\

\hline
22.5--23.0  & 0.856 $\pm$ 0.431      & 2.091 $\pm$ 0.536 \\
23.0--23.5  & 0.578 $\pm$ 0.250      & 1.235 $\pm$ 0.319 \\
23.5--24.0  & 0.125 $\pm$ 0.089      & 0.657 $\pm$ 0.198 \\
24.0--25.0  & 0.177 $\pm$ 0.072      & 0.532 $\pm$ 0.176 \\
25.0--26.0  & 0.089 $\pm$ 0.051      & 0.178 $\pm$ 0.102 \\
\hline
\label{tab5}
\end{tabular}
\begin{list}{}{}
\item[$^{\mathrm{a}}$] The bin widths are different since they  were chosen so 
as to always have at least two objects per bin. The differential RLF values 
have been normalized to the interval $\Delta$\,log P $=$ 0.5.
\end{list}
\end{center}
\end{table}

%------------------------------Table 6 --------------------------------------
\tabcolsep 0.8cm
\begin{table}[htb]
\begin{center}
\caption{The NEP Differential and Integral RLFs (Statistical Contaminant 
Method)}
\begin{tabular}{c|cc}
\\
\hline\hline
          & Differential     & Integral  \\
\hline
Log P$_{1.4}$ $^{\mathrm{a}}$ & N/cluster        & N/cluster \\
\hline
22.5--23.0  & 1.303 $\pm$ 0.682      & 2.532 $\pm$ 0.751 \\
23.0--23.5  & 0.392 $\pm$ 0.220      & 1.229 $\pm$ 0.314 \\
23.5--24.0  & 0.276 $\pm$ 0.131      & 0.837 $\pm$ 0.224 \\
24.0--25.0  & 0.163 $\pm$ 0.069      & 0.561 $\pm$ 0.181 \\
25.0--25.5  & 0.117 $\pm$ 0.083      & 0.235 $\pm$ 0.117 \\
25.5--26.0  & 0.118 $\pm$ 0.083      & 0.118 $\pm$ 0.083 \\

\hline
\label{tab6}
\end{tabular}
\begin{list}{}{}
\item[$^{\mathrm{a}}$]  The bin widths are different since they  were chosen so
as to always have at least two objects per bin. The differential RLF values
have been normalized to the interval $\Delta$\,log P $=$ 0.5.
\end{list}
\end{center}
\end{table}
%----------------------------------------------------------------------------

%------------------------------Table 7 --------------------------------------
\tabcolsep 0.4cm
\begin{table}[htb]
\begin{center}
\caption{Parameters of the Distant and Local  RLFs}
\begin{tabular}{c|cccc}
\\
\hline 
\hline
                    &  NEP  &               &  Ledlow \& Owen &           \\
\hline
Log P$_{1.4}$  & Slope & Normalization & Slope & Normalization\\
\hline
22.5--26.0     & 0.42$\pm$ 0.10 & 0.45$\pm$0.12 & 0.24 $\pm$ 0.04 & 0.18$\pm$0.02\\
~~~~$\le$ ~24.8& 0.55$\pm$ 0.22 & 0.50$\pm$0.17 & 0.10 $\pm$ 0.05 & 0.16$\pm$0.04\\
\hline	       
\label{tab7}
\end{tabular}
\end{center}
\end{table}
%----------------------------------------------------------------------------

\clearpage
\appendix

%-----------------------------------------------------------------
\section {Data Analysis}
\subsection {Bandwidth Smearing}
\label{bs}  

The use in synthesis observations of finite bandwidths produces a sort 
of chromatic aberration in the images consisting in a radial source
smearing which worsens with increasing distance from the phase center. 
The uncorrected point--source response is the convolution of the 
unsmeared restoring beam ($\theta_b$) with a radially extended
bandwidth--smearing ``beam''. Bandwidth-smeared images can be treated
as if obtained with a position--dependent 
synthesized beam ($\theta_{\rm bws}$) with major axis given by 
eq.~\ref{b_smea} and  oriented roughly at the same position angle of the
source position with respect to the field center. The peak response of a 
point source declines while the total flux density is conserved:
i.e. $S_P/S_T$ decreases with increasing distance from the field center. 
The smeared beam is still approximately gaussian and, since variances add 
in the convolution, its width can be estimated from the relation:
\begin{equation}
\frac{S_P}{S_T}\sim\frac{\theta_b}{\theta_{\rm bws}} ~~~~{\rm or} ~~~~
\theta_{\rm bws}\sim\theta_b\times \left(\frac{S_P}{S_T}\right)^{-1}
\label{b_smea}
\end{equation}

\noindent
The net effect is a radial degradation in the resolution and sensitivity of 
the array. For the VLA  the dependence of bandwidth smearing on the source
angular distance from the phase center ($d$), passband width 
$(\bigtriangleup\nu)$, observing frequency is well approximated  by the 
equation \citep{con98}:

\begin{eqnarray}
\label{eqn:bs}
\frac{S_{P}}{S^0{_{P}}} = \frac{S_P}{S_T} =\frac{1}{\displaystyle{\sqrt{1+\frac{2ln2}{3}
\left( \frac{\Delta\nu}{\nu}\frac{d}{\theta_b}\right)^2}}}
\end{eqnarray}
 
\noindent
where  ${S_{P}}/{S^0{_{P}}}$ is  the attenuation (${\cal A}$) of the
source peak brightness with respect to the unsmeared ($d = 0$) source peak
value. 
%------------------------------------------------------------------
%Fig A1 
\begin{figure}
   \centering
   \includegraphics[bb=0 120 574 624, width=14cm]{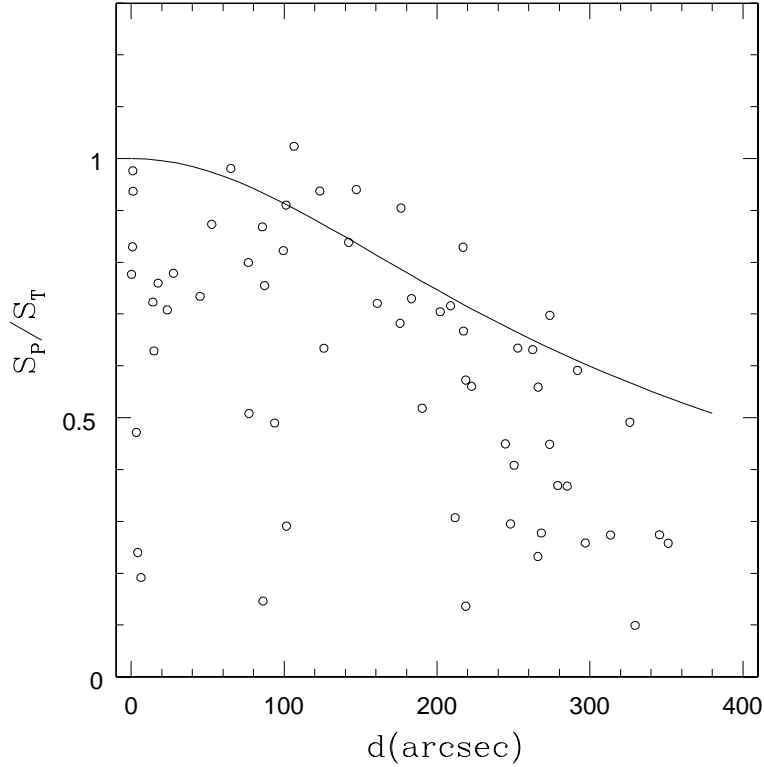}
   \vspace{-1cm}
   \caption{Ratio of the peak brightness $S_{P}$ to the integrated flux 
            density $S_T$ for radio sources with $S_{P} >$~$ 0.340$ \mJ 
            as a function of angular distance $d$ from the phase 
            center. The {\it solid line} represents  eq.~\ref{eqn:bs}.}
\label{figA1}
\end{figure}  
%------------------------------------------------------------------

\medskip
\noindent
We plot in  Fig.~\ref{figA1} the curve of eq.~\ref{eqn:bs} with 
$\nu = 1.4$ GHz, $<\theta_b> \simeq 4.5$\arcsec and $\Delta\nu = 42$ MHz.
The value adopted for $\theta_b$ is the average beam size (roughly
round) of the experiment. The value $\Delta\nu=$42 MHz was chosen 
since it is the best-fit effective bandwidth obtained by the NVSS 
survey data \citep{con98}. The circles in Fig.~\ref{figA1} indicate the 
$S_{P}/S_T$ ratio as a function of the angular distance $d$ for all the 
sources with $S_{P} \geq 0.340$ \mJ  ($\sim10\sigma$) within 350 arcsec
from the field center. The curve describes well, within the errors,
the upper envelope to the data. The curve  roughly indicates the location 
of point-like sources, demonstrating that it is  a good approximation for 
the peak  reduction due to bandwidth  smearing. 

\medskip
\noindent
Throughout this paper, radio sources are considered point-like only 
when the observed major axis is similar in size and orientation to 
$\theta_{\rm bws}$ (see Section~\ref{sample}).

%______________________________________________________________
\subsection {Optimization of Flux Densities}
\label{OPTS}

\cite{con98} discuss the effects of the noise on the determination of
the flux density.  They demonstrate  that the peak amplitude $S_P$ must be
corrected for a general statistical bias $\Delta S \approx \sigma^2/S_P$
(actually marginal for our data) caused by noise fluctuations which push 
the peak in the direction of the local noise increase. Then the corrected 
peak amplitude to use is: 
\begin{eqnarray}
S_{peak}  = S_P  -  \sigma^2/S_P
\label{Sbias}
\end{eqnarray}

\noindent
We further  correct for  bandwidth smearing according to 
eq.~\ref{eqn:bs}:
\begin{equation}
S_{Pcorr} = \frac{S_{peak}}{\cal A} = S_{peak} \times {\cal C}_{\rm
  bws}
\label{Pcorr}
\end{equation}

\noindent
In addition \cite{con97} and  \cite{con98} have shown that the optimization 
of the formal flux densities determined by  a bidimensional Gaussian fit, 
depends on whether the source apparent size is significantly larger than the
effective beam $ \theta_b \times \theta_{\rm bws}$ (see eq.~\ref{b_smea}). 

\noindent
For convenience of the reader we repeat below the formulae to be used. The 
optimum  flux densities have been obtained  by choosing one of the  following 
three  possibilities:
\begin{enumerate}
\item If a source is significantly resolved in two dimensions $(r)$
no  optimization is needed because of resolution thus  the peak
brightness is given simply  by  eq.~\ref{Pcorr}, and the
integrated flux density is: 
\[S_I ~ = ~ S_{Pcorr}~(\theta_M\theta_m)/(\theta_b\theta_{\rm bws})~=~
S_{peak}(\theta_M\theta_m/\theta_b^2)\]
\item If only the major axis is significantly resolved $(ru)$ the best 
estimates of the source peak amplitude and integrated flux density are:
\[S^{\dagger} ~ \equiv S_{Pcorr}(\theta_m/\theta_b)^{1/2}; ~~~~~ 
S_I = S^\dagger(\theta_M/\theta_{\rm bws})=S_{peak}(\theta_M/\theta_b)(\theta_m/\theta_b)^{1/2}\]
\item If neither axis is significantly resolved $(u)$, the best
estimate for both the peak brightness and integrated flux density is:
\[S_I\equiv S^* \equiv S_{Pcorr}(\theta_M\theta_m/\theta_b\theta_{bws})^{1/2}
= S_{peak}(\theta_M\theta_m/\theta_b^2)^{1/2}{\cal C}_{bws}^{1/2}\]
\end{enumerate}

%______________________________________________________________
\subsection {Errors on Source Parameters}
\label{errors}

The fractional uncertainty on source parameters can be approximated by the 
quadratic sum of two terms: 1) the intensity-independent calibration 
uncertainty,  $\epsilon$, and 2) the  noise-like uncertainty, $1/\rho$, due 
to the map noise.  The a--priori calibration uncertainties are  
$\epsilon=0.03$ for flux densities and $\epsilon=0.02$ for diameters.

To estimate the noise-like error on those source parameters which are the 
result of two-dimensional elliptical Gaussian fits (i.e. flux density, 
position and size) we adopt the approach of \cite{con97} who, on the 
assumption of a Gaussian noise distribution, has formally derived, 
the appropriate expressions. Thus the total fractional errors for peak
brightnesses $\left(\sigma_{S_P}/S_P\right)$ and for the fitted major 
and minor axes FWHM 
$\left(\sigma_{\theta_M}/\theta_M, \sigma_{\theta_m}/\theta_m\right) $
are given by the same type of equation:

\begin{equation}
\sqrt{\frac{2}{\rho^2}+\epsilon^2}{\rm ~~~ with ~~~}
{\rho^2}  =  \frac{\theta_M\theta_m}{4{\theta_N^2}}\left[ 1+\left( \frac{\theta_N}{\theta_M}\right)^2\right]^{\alpha_M}\left[ 1+\left( \frac{\theta_N}{\theta_m}\right)^2\right]^{\alpha_m}\left(\frac{S_P}{\sigma}\right)^2
\label{eq3}
\end{equation}   
\noindent
where $\epsilon$ is the a--priori calibration uncertainty.

\noindent
Here  $\sigma$  is the image noise and  $\theta_N$ the FWHM of the Gaussian
correlation length of the image noise (assumed to be equal to $\theta_b$).
The exponents are empirically determined and are equal to:

\smallskip
\noindent
$\alpha_M = 5/2$ and $\alpha_m = 1/2$ for calculating $\sigma_M$,\\
$\alpha_M = 1/2$ and $\alpha_m = 5/2$ for calculating $\sigma_m$,\\
$\alpha_M = 3/2$ and $\alpha_m = 3/2$ for calculating $\sigma_{S_P}$.\\

\noindent
The uncertainties on the optimized flux densities (Section~\ref{OPTS}) 
according to \cite{con98} are then given, in order, by: 

\begin{enumerate}
\item [1.] For an $(r)$ source:
\[\left(\frac{\displaystyle{\sigma_{S_I}}}{\displaystyle{S_I}}\right)^2 
\approx \left(\frac{\displaystyle{\sigma_{S_{Pcorr}}}}{\displaystyle{S_{Pcorr}}}
\right)^2 + 
\frac{\displaystyle{\theta_N^2}}{\displaystyle{\theta_M\theta_m}}
\left[ \left(\frac{\displaystyle{\sigma_{\theta_M}}}
{\displaystyle{\theta_M}}\right)^2+\left(\frac{\displaystyle{\sigma_{\theta_m}}}
{{\displaystyle{\theta_m}}}\right)^2\right]
{\rm ~~~~and~~~~} 
\sigma_{S_{Pcorr}}^2 ~ \approx ~ \left(\epsilon S_{Pcorr}\right)^2 +
2\left(\frac{\displaystyle{S_{Pcorr}}}{\displaystyle{\rho}}\right)^2\]
\item [2.] For an $(ru)$ source:
\[\left(\frac{\displaystyle{\sigma_{S_I}}}{\displaystyle{S_I}}\right)^2
\approx
\left(\frac{\displaystyle{\sigma_{S^\dagger}}}{\displaystyle{S^\dagger}}\right)^2 + 
\frac{\theta_N}{\theta_M} \left(\frac{\sigma_{\theta_M}}{\theta_M}\right)^2
{\rm ~~~~and~~~~}
\sigma_{S^{\dagger}}^2 \approx \left(\epsilon S^\dagger\right)^2 +
\frac{\displaystyle{3}}{\displaystyle{2}}
\left(\frac{S^\dagger}{\rho}\right)^2\]
\item [3.] For a $(u)$ source:
\[\sigma_{S_I}^2 \equiv \sigma_{S^*}^2 \approx \left(\epsilon  S^*\right)^2 + 
\left(\frac{\displaystyle{S^*}}{\displaystyle{\rho}}\right)^2 {\rm ~~~~or~~~~}
\left(\frac{\displaystyle{\sigma_{S_I}}}{\displaystyle{S_I}}\right)^2 
\approx \epsilon^2 +
\left(\frac{\displaystyle{1}}{\displaystyle{\rho}}\right)^2\]
\end{enumerate}

\medskip
\noindent
Here $\theta_N\approx \theta_b$ and $S_{Pcorr}$ is from eq. \ref{Pcorr}.
Note that we use now $S_I$, instead of $S_T$ used Section~\ref{bs}, to 
indicate that the integrated flux density has been optimized acccording to 
points 1, 2, 3 in Section~\ref{OPTS}).

\smallskip
\noindent
For the sources whose parameters were obtained with {\em TVSTAT}
(Section~\ref{sample}) the  noise error on 
the peak brightness is the image noise. The noise error on the 
total flux density is obtained by multiplying the image noise by the 
square-root of the number of  independent beam areas across the source. 
This is a good  approximation of eq.~\ref{eq3} when $ \theta_M, 
\theta_m \gg \theta_N$.

\end{document}